\documentclass[tighten,times,twocolumn]{aastex701}

\graphicspath{{./}}

\usepackage{amsmath}
\usepackage{mathtools}

\newcommand{\lp}{\ensuremath{\left(}}        
\newcommand{\rp}{\ensuremath{\right)}}

\def\beq{\begin{equation}}
\def\eeq{\end{equation}}

\renewcommand{\vec}[1]{{\boldsymbol{\mathbf{#1}}}}
\newcommand{\mat}[1]{{\boldsymbol{\mathbf{#1}}}}

\newcommand{\vk}{\vec{k}}

\newcommand{\vx}{\vec{x}}

\newcommand{\mC}{\mat{C}}

\newcommand{\appropto}{\mathrel{\vcenter{
  \offinterlineskip\halign{\hfil$##$\cr
    \propto\cr\noalign{\kern2pt}\sim\cr\noalign{\kern-2pt}}}}}

\newcommand{\hMpc}{\;h^{-1}\:\mathrm{Mpc}}
\newcommand{\ihMpc}{\;h\:\mathrm{Mpc}^{-1}}

\newcommand\numberthis{\addtocounter{equation}{1}\tag{\theequation}}

\usepackage{hyperref}
\usepackage{xspace}
\usepackage{graphicx}
\usepackage{hyperref}
\usepackage{booktabs}

\usepackage{savesym}
\savesymbol{tablenum}
\usepackage{siunitx}
\restoresymbol{SIX}{tablenum}

\usepackage[capitalize,noabbrev]{cleveref}
\newcommand{\secref}[1]{Section~\ref{#1}}

\newcommand{\datestart}{2019 January 1\xspace}
\newcommand{\dateend}{2019 November 5\xspace}
\newcommand{\kB}{\ensuremath{k_{\rm B}}}

\newcommand{\HI}{\ensuremath{{\rm HI}}}
\newcommand{\sHI}{\ensuremath{{\scriptscriptstyle {\rm HI}}}}

\newcommand{\Tb}{\ensuremath{T_{\rm b}}}
\newcommand{\Tbar}{\ensuremath{\bar{T}_{\rm b}}}

\newcommand{\OmegaHI}{\ensuremath{\Omega_\sHI}}

\newcommand{\deltaHI}{\ensuremath{\delta_\sHI}}

\newcommand{\deltam}{\ensuremath{\delta_{\rm m}}}
\newcommand{\tcm}{21$\,$cm\xspace}  

\newcommand{\bHI}{\ensuremath{b_\sHI}}

\newcommand{\zeff}{\ensuremath{z_\text{eff}}}
\newcommand{\AHI}{{\ensuremath{\mathcal{A}_\sHI}}}
\newcommand{\AHIps}{{\ensuremath{\mathcal{A}_\sHI^2}}}

\newcommand{\alphaNL}{\ensuremath{\alpha_{\rm NL}}}
\newcommand{\PL}{\ensuremath{P_{\rm L}}}
\newcommand{\PNL}{\ensuremath{P_{\rm NL}}}
\newcommand{\alphaFoG}{\ensuremath{\alpha_{\rm FoG}}}
\newcommand{\DFoG}[1]{\ensuremath{D_{#1}^{\rm FoG}}}
\newcommand{\DFoGSN}[1]{\ensuremath{D_{#1}^{\rm FoG,SN}}}

\newcommand{\fmu}{\langle f \mu^2\rangle}

\newcommand{\CNoise}{\ensuremath{\mC_{\rm N}}}
\newcommand{\AHIpsConstraintFullBand}{\AHIps = 3.55^{+0.96}_{-1.32}\text{(stat.)}\pm0.61\text{(sys.)}} 

\newcommand{\OmegaHIMargConstraint}{\OmegaHI(z=1.16) \times 10^3 = 0.79^{+0.20}_{-0.16}\text{(stat.)} \pm 0.07 \text{(sys.)}} 
\newcommand{\OmegaHIMargConstraintNoPrior}{\OmegaHI(z=1.16) \times 10^3 = 0.22^{+1.00}_{-0.07}\text{(stat.)} \pm 0.02 \text{(sys.)}} 
\newcommand{\OmegaHIMargConstraintQSOStack}{\OmegaHI(z=1.16) \times 10^3 = 0.85^{+1.18}_{-0.43}\text{(stat.)} }

\newcommand{\bHIMargConstraint}{\bHI(z=1.16) = 1.99^{+1.22}_{-0.74}\text{(stat)}
\pm 0.23 \text{(sys)}} 

\newcommand{\TNGOneHundredAmp}{2.38 \pm 0.18\text{(stat.)}\pm 0.40\text{(sys.)}} 
\newcommand{\TNGThreeHundredAmp}{3.97 \pm 0.30\text{(stat.)}\pm 0.67\text{(sys.)}} 

\newcommand{\TNGOneHundredNSigmas}{3.1\sigma} 
\newcommand{\TNGThreeHundredNSigmas}{4.0\sigma} 

\shorttitle{Interpretation of CHIME 21cm Auto Power Spectrum}

\begin{document}

\title{Interpretation of 21 cm Auto Power Spectrum Measurement at $z\sim 1$ by the Canadian Hydrogen Intensity Mapping Experiment}

\newcommand{\UBC}{Department of Physics and Astronomy, University of British Columbia, Vancouver, BC, Canada}
\newcommand{\MITP} {Department of Physics, Massachusetts Institute of Technology, Cambridge, MA, USA}
\newcommand{\MITK} {MIT Kavli Institute for Astrophysics and Space Research, Massachusetts Institute of Technology, Cambridge, MA, USA}
\newcommand{\TRU}{Department of Physical Sciences, Thompson Rivers University, Kamloops, BC, Canada}
\newcommand{\PI}{Perimeter Institute for Theoretical Physics, Waterloo, ON, Canada}
\newcommand{\DRAO}{Dominion Radio Astrophysical Observatory, Herzberg Astronomy \& Astrophysics Research Centre, National Research Council Canada, Penticton, BC, Canada}
\newcommand{\UBCO}{Department of Computer Science, Math, Physics, and Statistics, University of British Columbia-Okanagan, Kelowna, BC, Canada}
\newcommand{\McGill}{Department of Physics, McGill University, Montréal, QC, Canada}
\newcommand{\MU}{\McGill}
\newcommand{\TSI}{Trottier Space Institute, McGill University, 3550 rue University, Montréal, QC H3A 2A7, Canada}
\newcommand{\UofTastro}{David A.\ Dunlap Department of Astronomy \& Astrophysics, University of Toronto, Toronto, ON, Canada}
\newcommand{\UTA}{\UofTastro}
\newcommand{\UofTphys}{Department of Physics, University of Toronto, Toronto, ON, Canada}
\newcommand{\WVU} {Department of Computer Science and Electrical Engineering, West Virginia University, Morgantown WV, USA}
\newcommand{\WVUA} {Department of Physics and Astronomy, West Virginia University, Morgantown, WV, USA}
\newcommand{\WVUGWAC} {Center for Gravitational Waves and Cosmology, West Virginia University, Morgantown, WV, USA}
\newcommand{\WVUGC} {Center for Gravitational Waves and Cosmology, West Virginia University, Morgantown, WV, USA}
\newcommand{\Yale}{Department of Physics, Yale University, New Haven, CT, USA}
\newcommand{\YUP}{\Yale}
\newcommand{\YaleA}{Department of Astronomy, Yale University, New Haven, CT, USA}
\newcommand{\Dunlap}{Dunlap Institute for Astronomy and Astrophysics, University of Toronto, Toronto, ON, Canada}
\newcommand{\DIAA}{\Dunlap}
\newcommand{\RRI}{Raman Research Institute, Sadashivanagar,   Bengaluru, India}
\newcommand{\ASIAA}{Institute of Astronomy and Astrophysics, Academia Sinica, Taipei, Taiwan}
\newcommand{\CITA}{Canadian Institute for Theoretical Astrophysics, Toronto, ON, Canada}
\newcommand{\CIFAR}{Canadian Institute for Advanced Research,  Toronto, ON, Canada}
\newcommand{\WVUphysastro}{Department of Physics and Astronomy, West Virginia University, Morgantown, WV, USA}
\newcommand{\KIPAC}{Kavli Institute for Particle Astrophysics and Cosmology, Stanford, CA 94305, USA}
\newcommand{\SLAC}{SLAC National Accelerator Laboratory; Menlo Park, CA 94025; USA}
\newcommand{\ASU}{Department of Physics, Arizona State University, Tempe, AZ 85287, USA}
\newcommand{\RUG}{Kapteyn Astronomical Institute, University of Groningen, PO Box 800, 9700 AV Groningen, The Netherlands}
\newcommand{\ASTRON}{ASTRON, The Netherlands Institute for Radio Astronomy, Oude Hoogeveensedijk 4, Dwingeloo, 7991 PD, The Netherlands}
\newcommand{\UWO}{Department of Physics \& Astronomy, University of Western Ontario, 1151 Richmond Street, London, ON, N6A 3K7, Canada}


\shortauthors{CHIME Collaboration}
\collaboration{100}{The CHIME Collaboration:}

\author[0000-0001-6523-9029]{Mandana Amiri}
    \email{}
    \affiliation{\UBC}

\author[0000-0003-3772-2798]{Kevin Bandura}
    \email{}
    \affiliation{\WVU}
    \affiliation{\WVUGC}

\author[0000-0002-7758-9859]{Arnab Chakraborty}
    \email{}
    \affiliation{\MU}
    \affiliation{\TSI}

\author[0009-0003-8959-1918]{Zhuo Yu Brian Chu}
    \email{}
    \affiliation{\UBC}

\author[0000-0001-7166-6422]{Matt Dobbs}
    \email{}
    \affiliation{\MU}
    \affiliation{\TSI}

\author[0000-0002-0190-2271]{Simon Foreman}
    \email{}
    \affiliation{\ASU}

\author[0000-0003-3986-954X]{Liam Gray}
    \email{}
    \affiliation{\UBC}

\author[0000-0002-1760-0868]{Mark Halpern}
    \email{}
    \affiliation{\UBC}

\author[0000-0002-4241-8320]{Gary Hinshaw}
    \email{}
    \affiliation{\UBC}

\author[0000-0003-4179-4073]{Albin Joseph}
    \email{}
    \affiliation{\ASU}

\author[0009-0004-2241-0550]{Nolan Kruger}
    \email{}
    \affiliation{\ASU}

\author[0000-0001-8064-6116]{Joshua MacEachern}
    \email{}
    \affiliation{\DRAO}

\author[0000-0002-4279-6946]{Kiyoshi W. Masui}
    \email{}
    \affiliation{\MITK}
    \affiliation{\MITP}

\author[0000-0002-0772-9326]{Juan Mena-Parra}
    \email{}
    \affiliation{\DIAA}
    \affiliation{\UTA}

\author[0000-0002-7333-5552]{Laura Newburgh}
    \email{}
    \affiliation{\YUP}

\author[0000-0002-9516-3245]{Tristan Pinsonneault-Marotte}
    \email{}
    \affiliation{\KIPAC, SLAC}

\author[0000-0001-6967-7253]{Alex Reda}
    \email{}
    \affiliation{\YUP}

\author[0000-0001-6731-0351]{Shabbir Shaikh}
    \email{}
    \affiliation{\ASU}

\author[0000-0003-2631-6217]{Seth R. Siegel}
    \email{}
    \affiliation{\PI}
    \affiliation{\MU}
    \affiliation{\TSI}

\author[0009-0003-4114-1301]{Yukari Uchibori}
    \email{}
    \affiliation{\UBC}

\author[0000-0003-4535-9378]{Keith Vanderlinde}
    \email{}
    \affiliation{\UTA}
    \affiliation{\DIAA}

\author[0000-0002-1491-3738]{Haochen Wang}
    \email{}
    \affiliation{\MITP}
    \affiliation{\MITK}

\author[0000-0001-7314-9496]{Dallas Wulf}
    \email{}
    \affiliation{\MU}
    \affiliation{\TSI}

\correspondingauthor{Albin Joseph, Shabbir Shaikh}

\begin{abstract}
Observations with the Canadian Hydrogen Intensity Mapping Experiment (CHIME) have been used to measure the \tcm intensity mapping 
auto 
power spectrum, 
at $z\sim 1$, 
over a frequency range from  \SIrange{608.2}{707.8}{\mega\hertz} at wavenumbers $0.4\ihMpc \lesssim k \lesssim 1.5\ihMpc$.
In this paper, we present the results of two different approaches to interpreting this measurement.
In the first approach, we use a parametric power spectrum model to 
constrain an amplitude parameter, defined as $\AHIps \equiv 10^6 \OmegaHI^2(\bHI^2+\fmu)^2$, where $\OmegaHI$ is the cosmological density parameter for atomic hydrogen (\HI), $\bHI$ is the linear bias for {\HI}, and $\fmu$ incorporates the dominant large-scale impact of redshift-space distortions on the angle-averaged power spectrum.
Imposing an additional prior on either $\OmegaHI$ or $\bHI$, based on values in the literature, allows us to break the pairwise degeneracy between those two parameters.
%
In the second approach, we compare CHIME's measurement with predictions for the power spectrum of {\HI} from the IllustrisTNG simulations, finding that the measurement disagrees with the TNG100 run at $\TNGOneHundredNSigmas$ and the TNG300 run at $\TNGThreeHundredNSigmas$.
This disagreement is most likely attributable to the strength of nonlinear redshift-space clustering of {\HI} in the simulations, rather than the total abundance of {\HI}, and invites further investigation of the physical processes in the simulations that determine the behavior of {\HI} at nonlinear scales.
These results exemplify the ability of \tcm intensity mapping to provide astrophysical information using measurements at nonlinear scales.
\end{abstract}


\section{Introduction}

The technique of \emph{\tcm intensity mapping} is a promising avenue to map cosmic large-scale structure. This technique involves measuring the intensity of redshifted emission or absorption associated with hyperfine transitions in atomic hydrogen (\HI) and using the fluctuations in this intensity to infer the density of {\HI} at different locations. These measurements are being pursued across a wide range of redshifts, but in this work we are concerned with intensity mapping in the post-reionization epoch, specifically around redshift $z\sim 1$.

In this epoch, the vast majority of the {\HI} is co-located with galaxies, such that, at the largest scales, \tcm intensity maps provide a coarse-grained view of the {\HI}-mass-weighted distribution of galaxies. At these scales, intensity maps can therefore be interpreted with the same perturbative theoretical framework used to describe the distribution of optically-selected galaxies and other biased tracers of large-scale structure \citep[e.g.][]{villaescusa-navarro2018,castorina2019-growth,karagiannis2020-nGforecasts,pourtsidou2023-forecasts,obuljen2023-fieldlevelHI,foreman2024-HIstoch}. This implies that such maps can provide similarly rich cosmological information, by measuring baryon acoustic oscillations, linear growth, the full shape of the power spectrum, and non-Gaussian summary statistics.

At smaller spatial scales where clustering becomes nonlinear, the perturbative framework is no longer valid and first-principles modelling is much more difficult. However, measurements at these scales probe the distribution of {\HI} in the environments of dark matter halos and galaxies, and also probe the {\HI} velocity field via redshift-space distortions. Elucidating these properties of {\HI} can aid in understanding the baryon cycle that moves gas between the intergalactic medium, circumgalactic medium, and galactic environment, with important consequences for our understanding of galaxy evolution \citep[e.g.][]{peroux2020-baryoncycle,koribalski2020-wallaby,2020walter-baryons}. In turn, this understanding can aid in the interpretation of other cosmological observations, such as cosmic shear \citep{chisari2019-baryonicfeedback} and the Sunyaev-Zeldovich effect \citep{mroczkowski2019-sz}, that are sensitive to so-called ``baryonic effects" (e.g.\ feedback from active galactic nuclei) at nonlinear scales.

\citet{chime-auto-paper} has presented the first measurement of the \tcm auto power spectrum at $z\sim 1$, using data from the Canadian Hydrogen Intensity Mapping Experiment (CHIME; \citealt{CHIMEoverview}). That work described a series of data-processing improvements that have been implemented since CHIME's earlier \tcm cross-correlation detections \citep{chimestacking,chime-lymanalpha}, and these improvements have enabled the \tcm power spectrum to be measured with signal to noise ratio $12.4$ in the frequency band from \SIrange{608.2}{707.8}{\mega\hertz}. That work also described a comprehensive suite of validation tests that demonstrated the consistency of the measured power spectrum between several analysis choices and splits of the data, as well as confirming that the inferred amplitude of the power spectrum is consistent with the stacking analysis of \citet{chimestacking}. Importantly, this measurement was made in the wavenumber range $0.4\ihMpc \lesssim k \lesssim 1.5\ihMpc$, which is in the nonlinear regime discussed above.\footnote{See \citet{paul2023-21cmauto} and \citet{townsend-lowz-auto2026} for other reported detections of the \tcm auto power spectrum, also at nonlinear scales.}

In this work, we present the results of two different approaches to interpreting the power spectrum measurement from \citet{chime-auto-paper}. 
In both approaches, our primary goal is to focus on uncertainties in the astrophysical processes that generate the signal, as opposed to uncertainties in the underlying cosmological model.

In the first approach, we fit a parametric model for the \tcm power spectrum. 
This model contains two parameters that mostly affect the power spectrum amplitude: the cosmological density parameter for {\HI}, $\OmegaHI(z)$ (defined as the ratio of the mean comoving density of {\HI} at redshift~$z$ to the critical density at $z=0$), and the linear bias of {\HI}, $\bHI(z)$, which captures the linear part of the relationship between the clustering of {\HI} and the clustering of all matter. 

In addition, the model contains two parameters that affect the shape of the power spectrum. However, these parameters also significantly affect the power spectrum amplitude at nonlinear scales. At the signal to noise ratio of the present measurement, the observed bandpowers primarily constrain the overall amplitude of the power spectrum, with less sensitivity to variations in shape.
This leads to strong degeneracies between $\OmegaHI$, $\bHI$, and the shape parameters, as well as susceptibility to prior-volume and projection effects, which we discuss in detail. We also discuss several tests we have done to validate our choice of model and parameter priors.

While $\OmegaHI$ and $\bHI$ have well-defined physical meanings, the shape parameters are purely phenomenological: they are chosen to ensure the model is flexible enough to encompass a reasonable range of power spectrum shapes, but their values do not have a robust physical interpretation. Therefore, in our Bayesian analysis, we choose to marginalize over these values. Meanwhille, $\OmegaHI$ and $\bHI$ are highly degenerate, so we choose to constrain a single effective amplitude parameter, defined by $\AHIps = 10^6 \OmegaHI^2(\bHI^2+\fmu)^2$, at the mean redshift of the measurement ($z=1.16$). In this expression, the $\fmu$ term incorporates the appropriate angular average of the linear (Kaiser) contribution from redshift-space distortions, at the scales of our measurement.

Our marginalized constraint on this amplitude is
$\AHIpsConstraintFullBand$, where the systematic uncertainty budget combines several potential sources of error, which we quantify in detail.
We also explore the constraints that can individually be placed on~$\OmegaHI$ and~$\bHI$, if an additional prior is placed on one of these parameters to break the pairwise degeneracy between them. 
If we assume a Gaussian prior on $\bHI(z=1.16)$ with mean $1.56$ and standard deviation that is 20\% of this value, based on a compilation of simulation measurements in the literature, then we obtain $\OmegaHIMargConstraint$. 
If, on the other hand, we assume a Gaussian prior on $\OmegaHI(z=1.16)$ with mean $6.3\times 10^{-4}$ and standard deviation that is 25\% of this value, based on previous estimates from {\HI} stacking, intensity mapping cross-correlations, and damped Lyman-$\alpha$ systems, we obtain $\bHIMargConstraint$.

In the second approach to interpreting the measured power spectrum, we assess the consistency of the measurement with {\HI} power spectra measured from the TNG100 and TNG300 runs of the IllustrisTNG suite of cosmological hydrodynamical simulations \citep{TNGa,TNGb,TNGc,TNGd,TNGe}. 
In detail, we measure redshift-space power spectra from simulation snapshots at $z=1$, and use them as inputs for the CHIME simulation pipeline, which generates multifrequency sky maps and then propagates these maps into visibilities. We process these visibilities with the same power spectrum pipeline as the data; as a result, the impact of each step in the pipeline (filtering, masking, and so on) on the input power spectrum is self-consistently foward-modelled.

We compare these simulation-derived IllustrisTNG predictions with the power spectrum from \citet{chime-auto-paper} by scaling the prediction with a free amplitude and constraining the value of this amplitude. For TNG100, we find $\TNGOneHundredAmp$ for this amplitude, while for TNG300, we find $\TNGThreeHundredAmp$. This implies that the predictions are discrepant with the measured power spectrum at $\TNGOneHundredNSigmas$ and $\TNGThreeHundredNSigmas$, respectively. If each TNG prediction is rescaled by its best-fit amplitude, the resulting curves provide a good fit to the measured bandpowers, so the discrepancy is primarily due to the simulations having a low amplitude of {\HI} clustering compared to the data, at the scales where the measurement was made.
We argue that this discrepancy is unlikely attributable to the total amount of {\HI} in the simulations, as indicated by $\OmegaHI$; instead, it indicates a mismatch between the amount of redshift-space {\HI} clustering observed by CHIME and predicted by the simulations.

This paper is organized as follows:
\begin{itemize}
\item In \secref{sec:data}, we briefly summarize the dataset and processing pipeline used for the power spectrum measurements in \citet{chime-auto-paper}.
\item In \secref{sec:modelling}, we present our parametric power spectrum model, along with our method for using simulations to forward-model the transfer function associated with the telescope and power spectrum measurement pipeline.
\item In \secref{sec:simulations}, we describe our method for generating simulated datasets and power spectrum measurements.
\item In \secref{sec:posterior_estimation}, we discuss the details of our parameter estimation procedure, including our choices of parameter priors, the definition of the power spectrum amplitude parameter $\AHIps$, and the tests we have performed to validate our parametric model and analysis choices.
\item In \secref{sec:parameter_constraints}, we present posteriors from a Bayesian analysis using our power spectrum model, describe how we estimate the systematic uncertainty in our inference of $\AHIps$, discuss the impact of different priors, present an additional validation test using the IllustrisTNG simulations, and show separate constraints on~$\OmegaHI$ and~$\bHI$.
\item In \secref{sec:tng}, we compare the measured power spectrum with predictions from IllustrisTNG.
\item In \secref{sec:conclusions}, we conclude and discuss possible future directions for interpretation.
\end{itemize}
Appendices~\ref{app:model_details} and~\ref{app:sim_details}  provide further details of our pipelines for evaluating our power spectrum model and generating simulations.
Appendix~\ref{app:tng} describes our procedure for propagating {\HI} power spectra from IllustrisTNG through our simulation and analysis pipelines.

In this work, we use a $\Lambda$CDM cosmological model with parameter values from the final {\em Planck} data release (specifically, the ``TT,TE,EE+lowE+lensing+BAO" parameters from Table 2 of \citealt{planck2020}).

\section{Data and Processing}
\label{sec:data}

In this section, we describe the key properties of the dataset and processing pipeline that resulted in the power spectrum measurement in \citet{chime-auto-paper}. We refer the reader to that work for further details.

The nighttime portions of $94$ sidereal days from the period between \datestart and \dateend were selected for analysis based on several data-quality diagnostics. The analysis was further restricted to a frequency band of $256$ channels from \SIrange{608.2}{707.8}{\mega\hertz} ($1.34 > z > 1.01$ in \tcm redshift), chosen to avoid wide gaps in frequency coverage due to persistent radio-frequency interference (RFI) and spatial aliasing in the relevant portion of the sky.

RFI was masked in each sidereal day, using algorithms that (1) compared the variance of the visibilities to expectations from the radiometer equation, and (2) searched for excursions after applying a high-pass fringe-rate filter or high-pass delay filter. The visibilities were beamformed in the north-south direction, and a DAYENU \citep{ewall-wice2021} foreground filter was applied to remove slowly-varying spectral components corresponding to delays less than $200\,{\rm ns}$. The HyFoReS algorithm \citep{hyfores_2022,hyfores_2025a,hyfores_2025b} was applied to correct for bandpass-gain fluctuations, and a second stage of RFI flagging was performed. An estimate of the noise cross-talk level was removed from each day, the daily timestreams were rebinned onto a fixed grid in local Earth rotation angle, and these timestreams were averaged into two sets of ``stacked" visibilities, corresponding to even- and odd-indexed days.

The visibilities were transformed into a ``ringmap," which is effectively a dirty map equally-sampled in right ascension and the sine of the zenith angle along the local meridian, filtered to only contain information consistent with the telescope's response to the sky. Separate maps for co-polar baselines (XX, YY) were combined into a pseudo-Stokes-$I$ map. Several narrow-band spatially-local negative excursions were identified in the map pixel values; these are consistent with the properties of \tcm absorption systems, and were masked at this point.

The ringmaps were then Fourier-transformed in frequency, with a Wiener filter that accounts for the filtering and masking that has been applied. Further spatial masks were applied to excise regions associated with bright point sources and the galactic plane, and a flat-sky Fourier transform from angular space to $(u,v)$ space was carried out. Finally, a cross power spectrum between the even- and odd-day maps was computed to eliminate noise bias, with a binning scheme defined in terms of comoving cosmological wavenumbers $k$. The result is an auto power spectrum measured in $8$ bandpowers within $0.4\ihMpc \lesssim k \lesssim 1.5\ihMpc$. Over the same~$k$ range, power spectra were also measured for two sub-bands, from \SIrange{608.2}{658.2}{\mega\hertz} ($1.34 > z > 1.16$) and \SIrange{658.2}{707.8}{\mega\hertz} ($1.16 > z > 1.01$).

The noise covariances associated with the measured power spectra were estimated from 1000 realizations of Gaussian noise consistent with fast-cadence noise estimates from the CHIME real-time pipeline, incorporating the effects of filtering, masking, and other pipeline stages described above.


\section{Modelling}
\label{sec:modelling}

\subsection{Cosmological \tcm power spectrum}
\label{sec:modelling:cosmological}

The \tcm power spectrum holds a large amount of information about the background cosmological model, growth of matter perturbations, and astrophysics of {\HI} across a wide range of scales. Fully accessing this information requires a transfer function that captures the relationship between a theoretical power spectrum prediction and the observed power spectrum, accounting for properties of the telescope and analysis pipeline. A CHIME transfer function that is valid for a generic input power spectrum is currently under development, and future work will describe this transfer function and use it to obtain constraints on different aspects of cosmology and astrophysics.

In this work, we instead restrict ourselves to a specific form of the theoretical power spectrum that can be decomposed into terms that we call ``templates." For each template, we can compute its contribution to the observed power spectrum using simulations, and assemble the results into an expression whose model parameters can be varied efficiently enough for Markov Chain Monte Carlo sampling of the joint posterior of these parameters. This approach thereby avoids the need for a fully general transfer function.

Our power spectrum model contains ingredients that are physically motivated, but our modelling of nonlinear scales is approximate and contains parameters that are purely phenomenological.\footnote{In this sense, our nonlinear parameters can be viewed similarly to the parameters that capture the broadband shape of the galaxy power spectrum in analyses of baryon acoustic oscillations \citep[e.g.][]{anderson2014-bao,beutler2017-bao,bautista2021-bao,chen2024-bao}.}
Our goal is that our model is flexible enough to span the plausible range of nonlinear behavior of the \tcm power spectrum, such that we can marginalize over the associated parameters to obtain conservative constraints on larger-scale quantities (namely, the mean density and linear bias of {\HI}). A similar philosophy was adopted in the first cross-correlation between CHIME and eBOSS~\citep{chimestacking}.

Our model takes the following form:
\begin{align*}
P_{21}(k, \mu; z) 
	&= \Tbar(z)^2
	\left[ \bHI(z) + f(z) \mu^2 \right]^2 \\
&\quad \times
	P_{\rm m}(k, z)
	\DFoG{\sHI}(k\mu, z; \alphaFoG)^2\ ,
\numberthis
\label{eq:P21-theory}
\end{align*}
where
\begin{align*}
P_{\rm m}(k, z) 
	&= \left[ \frac{D^+(z)}{D^+(z_{\rm fid})} \right]^2
	\bigg[ \PL(k, z_{\rm fid})  \\
&\qquad+ \alphaNL \left\{ 
		\PNL(k, z_{\rm fid}) - \PL(k, z_{\rm fid}) 
	\right\} \bigg]\ .
\numberthis
\label{eq:Pm-model}
\end{align*}
We describe the ingredients for this model below.

\paragraph{Mean \tcm brightness temperature $\Tbar(z)$} 
The \tcm brightness temperature $\Tb(\vx, z)$ is equal to the {\HI} overdensity $\deltaHI(\vx, z)$ multiplied by the mean brightness temperature $\Tbar(z)$ (e.g.~\citealt{bull2015}), so the \tcm power spectrum is proportional to~$\Tbar(z)^2$. The mean brightness temperature is related to the mean $\HI$ density parameter $\OmegaHI(z)$ via \citep{chimestacking}
\beq
\Tbar(z) \approx 191.06 \left[ h \frac{H_0}{H(z)} \, \OmegaHI(z) \, (1 + z)^2 \right] \: \si{\milli\kelvin} \; .
\label{eq:Tbarcompact}
\eeq

The signal to noise ratio and redshift range of our power spectrum measurement are insufficient to constrain the redshift evolution of $\OmegaHI(z)$.
Therefore, we use a model that rescales a fiducial form for $\OmegaHI(z)$ by an amplitude parameter $\alpha_\Omega$:
\beq
\OmegaHI(z) = \alpha_\Omega\, \OmegaHI^{\rm (fid)}(z)\ .
\label{eq:OmegaHImodel}
\eeq
For $\OmegaHI^{\rm (fid)}(z)$, we use the fitting function from \citet{crighton2015}:
\beq
\OmegaHI^{\rm (fid)}(z) = 4\times 10^{-4} (1+z)^{0.6}\ .
\label{eq:OmegaHIfid}
\eeq
We will quote our main results in terms of the {\HI} density parameter at the effective redshift of our measurement, which, for brevity, we denote by $\OmegaHI$ instead of $\OmegaHI(\zeff)$:
\beq
\OmegaHI \equiv \alpha_\Omega\, \OmegaHI^{\rm (fid)}(\zeff)
	= \alpha_\Omega \times 6.35 \times 10^{-4}\ .
\label{eq:OmegaHIfitpar}
\eeq

\paragraph{Linear bias and Kaiser term $\left[ \bHI(z) + f(z) \mu^2 \right]$} We assume a scale-independent linear bias $\bHI(z)$ that relates the densities of {\HI} and matter on large scales. We account for the leading-order effect of redshift-space distortions on large scales by including the Kaiser term $f(z)\mu^2$, where $f(z)$ is the logarithmic growth rate of matter perturbations \citep{kaiser1987}.
Similarly to $\OmegaHI(z)$, we use a model that rescales a fiducial form for $\bHI(z)$ by an amplitude parameter $\alpha_b$:
\beq
\bHI(z) = \alpha_b\, \bHI^{\rm (fid)}(z)\ .
\label{eq:bHI-model}
\eeq
We use the same fiducial bias function as \citet{chimestacking}, which was fit to
 {\HI} bias measurements from the TNG100 run of IllustrisTNG \citep{villaescusa-navarro2018}:
\begin{align*}
\bHI^{\rm (fid)}(z) &= 1.489 + 0.460(z-1) - 0.118(z-1)^2 \\
&\quad+ 0.0678(z-1)^3 - 0.0128(z-1)^4 \\
&\quad+ 0.0009 (z-1)^5\ .
\numberthis
\label{eq:bHI-fitfunc}
\end{align*}
We do not fit for $f(z)$, instead keeping it fixed by our fiducial cosmology. Instead of quoting results in terms of $\alpha_b$, we will use the {\HI} bias at the measurement's effective redshift, writing $\bHI$ as an abbreviation of $\bHI(\zeff)$:
\beq
\bHI \equiv \alpha_b\, \bHI^{\rm (fid)}(\zeff) = \alpha_b \times 1.56\ .
\label{eq:bHI-fitpar}
\eeq

\begin{deluxetable*}{Lcl}[t]
    \tablecaption{Summary of free parameters in \tcm power spectrum model used in this work.\label{tab:params}}
    \tablecolumns{3}
    \tablewidth{\linewidth}
    \tablehead{
        \colhead{Parameter} &
        \colhead{Defining equations} &
        \colhead{Description} 
    }
    \startdata
        \hspace{4mm} \OmegaHI   
        		& \eqref{eq:P21-theory}, \eqref{eq:Tbarcompact}, \eqref{eq:OmegaHImodel}, \eqref{eq:OmegaHIfid}, \eqref{eq:OmegaHIfitpar}
		& \parbox[t]{0.6\textwidth}{Mean {\HI} density at effective redshift of measurement} \\
        \hspace{4mm} \bHI           
        		&  \eqref{eq:P21-theory}, 
		\eqref{eq:bHI-model}, 
                \eqref{eq:bHI-fitfunc},
                \eqref{eq:bHI-fitpar}
		& \parbox[t]{0.6\textwidth}{Linear {\HI} bias at effective redshift of measurement} \\
        \hspace{4mm} \alphaNL       
        		& \eqref{eq:Pm-model} 
		& \parbox[t]{0.6\textwidth}{Nonlinearity parameter, interpolating between linear and nonlinear matter power spectrum models}  \\
        \hspace{4mm} \alphaFoG
        		& \eqref{eq:P21-theory}, \eqref{eq:DFoGk}, \eqref{eq:sigmaFoGfid}
		& \parbox[t]{0.6\textwidth}{Finger-of-God parameter, multiplying fiducial model for redshift-dependent damping scale} \\
    \enddata
\end{deluxetable*}

\paragraph{Matter power spectrum $P_{\rm m}(k, z)$} 
We assume that the scale-dependence of the \tcm power spectrum factorizes into two contributions: the matter power spectrum $P_{\rm m}(k, z)$ and a damping factor $\DFoG{\sHI}(k\mu; z)^2$ arising from nonlinear redshift-space distortions, described below. We take the matter power spectrum to have the form in Eq.~\eqref{eq:Pm-model}: a weighted sum of linear and nonlinear models, with a parameter $\alphaNL$ that interpolates between pure-linear ($\alphaNL=0$) and pure-nonlinear ($\alphaNL=1$) cases\footnote{\citet{amon2022-shear} and \citet{preston2024-shear} have used a similar parameterization to describe the range of baryonic effects on the matter power spectrum in the context of cosmic shear.}, evaluated at $z_{\rm fid}=1$ and rescaled to arbitrary redshift using the square of the linear growth factor $D^+(z)$. We use the \texttt{CAMB} package \citep{lewis1999} to compute these matter power spectra, using the model from \citet{mead2021} for the nonlinear spectrum.

\paragraph{Finger-of-God damping $\DFoG{\sHI}(k\mu, z; \alphaFoG)$} We incorporate redshift-space distortions from small-scale velocities (``Fingers of God"; \citealt{jackson1972}) with the square of a Lorentzian damping factor,
\begin{align*}
&D_\sHI^{\rm FoG}(k\mu, z; \alphaFoG) \\
&\qquad
	= \frac{1}{1 + \frac{1}{2} \left[
		 k\,  \mu \,\alphaFoG\,   \sigma_{\rm FoG}^{\rm (fid)}(z)
	\right]^2}\ ,
	\numberthis
\label{eq:DFoGk}
\end{align*}
which depends on a fiducial model for the redshift-dependent damping scale $\sigma_{\rm FoG}^{\rm (fid)}(z)$. Our fiducial model for this scale is based on transforming the simulation measurements from \cite{sarkar2019} into an effective scale for squared-Lorentzian damping of the {\HI} power spectrum; the results are well-described by \citep{chimestacking}
\beq
\frac{\sigma_{\rm FoG}^{\rm (fid)}(z)}{h^{-1} {\rm Mpc}} 
	= 1.93 - 1.48 (z-1) + 0.81 (z-1)^2\ .
\label{eq:sigmaFoGfid}
\eeq
\cref{eq:DFoGk} includes
an overall scaling factor $\alpha_{\rm FoG}$
that we take to be a free parameter.
The model from \cite{sarkar2019} has difficulties reproducing measurements of {\HI} Fingers of God in some simulations \citep{villaescusa-navarro2018}, and therefore the damping scale in this model should not be strictly interpreted as representing the physical velocity dispersion of {\HI}, but instead as a phenomenological ``nonlinear shape" parameter, similar to $\alphaNL$ above.

\phantom{whitespace here}

The uncertainties in the above parameters are much larger than uncertainties on cosmological parameters, so we keep the background cosmology fixed to the \textit{Planck} 2018 model \citep{planck2020}. Table~\ref{tab:params} contains a summary of the free parameters in our signal model.

\cite{osinga2025-HImodelling} has shown that some previous {\HI} power spectrum models in the literature are not sufficiently flexible to capture the behavior observed in hydrodynamical simulations.
We stress-test our model in a similar fashion in \secref{sec:parameter_constraints:tngmock}; we find
that the model provides unbiased constraints on the power spectrum amplitude 
when the nonlinear parameters are marginalized over and {\HI} predictions from IllustrisTNG are used as mock data.

A separate shot noise contribution is sometimes included in models for the \tcm power spectrum (e.g.\ \citealt{paul2023-21cmauto,padmanabhan2023-meerkatinterpretation,li2024-HIautomodel,osinga2025-HImodelling}). We find that the model described above is sufficiently flexible
 to provide a good fit both to our observed power spectrum and measurements from hydrodynamical simulations, given the signal to noise ratio of our observations.
  Furthermore, based on the strong parameter degeneracies discussed in \secref{sec:modelling:template} and \secref{sec:parameter_constraints:posteriors}, we expect that additional model parameters would be poorly constrained by our current measurement. Therefore, we do not include
shot noise in our model.

\subsection{Template-based model evaluation}
\label{sec:modelling:template}

Using the $\alpha$ parameters defined in \secref{sec:modelling:cosmological}, we can rewrite \cref{eq:P21-theory,eq:Pm-model} 
as a linear combination of six terms:
\begin{align*}
P_{21}(k,\mu,z)
	&= \alpha_\Omega^2 \alpha_b^2 (1-\alphaNL) 
	P_{21}^{\rm hh,L}(k,\mu,z; \alphaFoG) \\
&\quad
	+ \alpha_\Omega^2 \alpha_b^2 \alphaNL 
	P_{21}^{\rm hh,NL}(k,\mu,z; \alphaFoG) \\
&\quad
	+ \alpha_\Omega^2 \alpha_b (1-\alphaNL) 
	P_{21}^{\rm hv,L}(k,\mu,z; \alphaFoG) \\
&\quad
	+ \alpha_\Omega^2 \alpha_b \alphaNL 
	P_{21}^{\rm hv,NL}(k,\mu,z; \alphaFoG) \\
&\quad
	+ \alpha_\Omega^2  (1-\alphaNL) 
	P_{21}^{\rm vv,L}(k,\mu,z; \alphaFoG) \\
&\quad
	+ \alpha_\Omega^2 \alphaNL 
	P_{21}^{\rm vv,NL}(k,\mu,z; \alphaFoG) \ ,
	\numberthis
	\label{eq:P21theory_sum}
\end{align*}
where we have defined
\begin{align*}
P_{21}^{X,Y}(k,\mu,z; \alphaFoG) 
&= \Tbar^{\rm (fid)}(z)^2 F_X(\mu, z)  \\
&\quad\times \left[ \frac{D^+(z)}{D^+(z_{\rm fid})} \right]^2
	P_Y(k, z_{\rm fid}) \\
\numberthis
&\quad
	\times
	  \DFoG{\sHI}(k\mu, z; \alphaFoG)^2\ .
\label{eq:P21XY}
\end{align*}
In \cref{eq:P21XY}, $X = {\rm hh}, {\rm hv}, {\rm vv}$, where ``${\rm h}$" and ``${\rm v}$" refer to factors of the linear bias and Kaiser term; $Y={\rm L}, {\rm NL}$ indicates whether the linear or nonlinear matter power spectrum is used; and
\begin{align}
F_{\rm hh}(\mu, z) &= \bHI^{\rm (fid)}(z)^2\ , \\
F_{\rm hv}(\mu, z) &= 2 \bHI^{\rm (fid)}(z) f(z) \mu^2\ , \\
F_{\rm vv}(\mu, z) &= f(z)^2 \mu^4\ .
\end{align}

Importantly, the near-linearity of the analysis pipeline in \citet{chime-auto-paper} implies that the observed power spectrum can be modelled as a cosmological power spectrum multiplied by a linear transfer function. (See Appendix~\ref{app:model_details} for further discussion.)
Therefore, if we can apply this transfer function to each~$P_{21}^{X,Y}$ term in \cref{eq:P21theory_sum}, we can evaluate the full power spectrum prediction at arbitrary values of $\alpha_\Omega$,~$\alpha_b$, and~$\alphaNL$ by forming the appropriate linear combination of these terms, with the parameter-dependent prefactors in \cref{eq:P21theory_sum}.
Since the prediction is nonlinear in~$\alphaFoG$, that parameter must be handled separately for each term.

In Appendix~\ref{app:model_details}, we describe the details of this procedure, but we provide an outline here. We generate several simulated visibility datasets, and apply the power spectrum pipeline from \citet{chime-auto-paper} to each one. 
In each case, the output is an ``observed" power spectrum that represents the application of the pipeline's transfer function to the corresponding input power spectrum, without requiring knowledge of the transfer function itself.
We use input \tcm power spectra evaluated at 6 specially-chosen points in $(\alpha_\Omega, \alpha_b, \alphaNL)$ parameter space and at 20 different values of $\alphaFoG$, resulting in a set of 120 simulations. 
\secref{sec:simulations:method} describes how these simulations are performed.

For each evaluation of the power spectrum model, these pre-computed power spectra are combined as follows:
\begin{enumerate}
\item At each value of $\alphaFoG$, we form specific linear combinations of the 6 ``observed" power spectra (see Appendix~\ref{app:model_details:template_combinations}). Each linear combination corresponds to one of the terms in \cref{eq:P21theory_sum}. We refer to these terms with $P_{\rm obs}^Z(k;\alphaFoG)$ where $Z$ stands for one of the 6 superscripts in \cref{eq:P21theory_sum} (e.g.\ ``${\rm hh,L}$").
\item For each $k$ and $Z$, we interpolate between the simulation results at different $\alphaFoG$ values to obtain $P_{\rm obs}^Z(k;\alphaFoG)$ at the desired value of $\alphaFoG$ (see Appendix~\ref{app:model_details:fog}).
\item For the desired values of $\alpha_\Omega$, $\alpha_b$, and $\alphaNL$, we evaluate \cref{eq:P21theory_sum} with $P_{\rm obs}^Z(k;\alphaFoG)$ in place of $P_{21}^{Z}(k,\mu,z; \alphaFoG)$.
\end{enumerate}
This procedure enables fast evaluations of the model without requiring new simulations at each point in parameter space. In \secref{sec:parameter_constraints:systematics:template_accuracy}, we quantify the accuracy of this approach by comparing the model's predictions with a set of validation simulations. 

\begin{figure}[t]
   \centering \includegraphics[width=\linewidth, keepaspectratio, trim = 0 5 0 0]{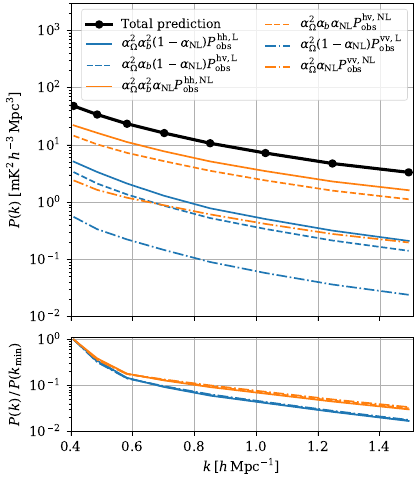}
    \caption{%
    \emph{Upper panel}: Template-based prediction for the observed \tcm power spectrum, evaluated at a representative point in parameter space, along with the 6 terms that are summed together to produce this prediction. Each term is computed by generating simulated visibilities with a specific input power spectrum, and applying our power spectrum measurement pipeline to these visibilities. 
    \emph{Lower panel}:  Each term is normalized by its value at $k_{\rm min}=0.41\ihMpc$ to highlight differences in shape, arising from whether the term involves the linear or nonlinear matter power spectrum, along with how many factors of the linear {\HI} bias and Kaiser factor ($f\mu^2$) are included.
    }
    \label{fig:templates}
\end{figure}

\cref{fig:templates} shows the 6 templates corresponding to the terms in \cref{eq:P21theory_sum} (i.e.\ the terms constructed in step 1 above), evaluated at 
a representative point in parameter space: $(\alpha_{\Omega}, \alpha_{b}, \alphaNL, \alphaFoG) = (1.0, 1.5, 0.75, 0.75)$.
The upper panel demonstrates that these terms typically have very different amplitudes, with the nonlinear ``${\rm hh}$" term (proportional to $\bHI^2$) providing the dominant contribution.\footnote{Note that, due to parameter degeneracies, the relative contributions of the different terms can vary in such a way that the sum is only changed by a small amount, so the hierarchy in \cref{fig:templates} is just an illustrative example.}

\begin{figure*}
   \centering \includegraphics[width=1.0\linewidth, keepaspectratio, trim = 0 5 0 0]{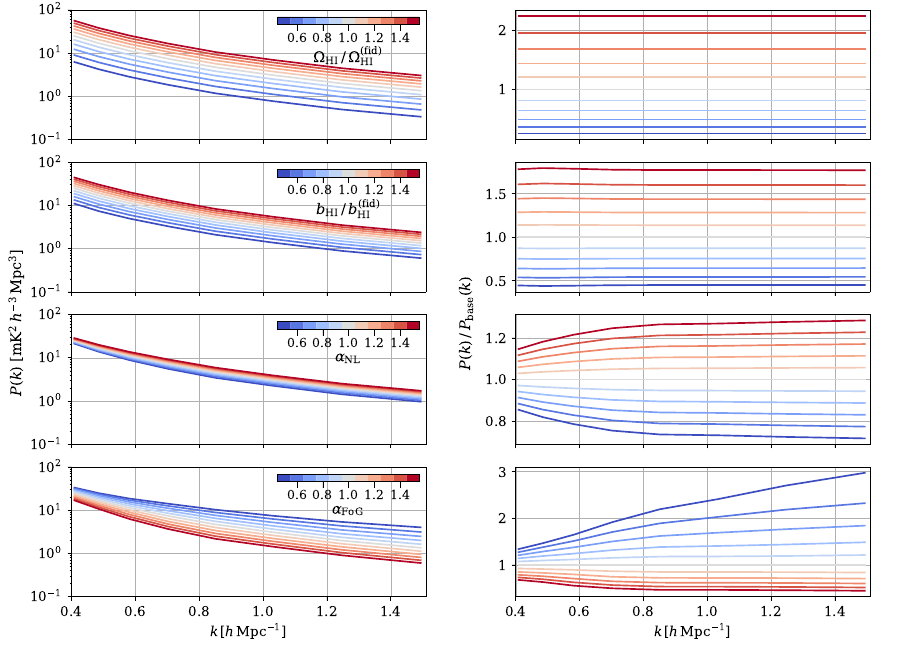}
    \caption{%
    \emph{Left panels}: Dependence of our \tcm power spectrum model on each parameter: {\HI} density parameter $\OmegaHI$, linear bias $\bHI$, nonlinearity parameter $\alphaNL$, and Finger-of-God damping parameter $\alphaFoG$. Each parameter significantly affects the overall amplitude of the power spectrum at the scales of interest in this work.
    \emph{Right panels}: Each power spectrum is divided by a ``base" version with $\OmegaHI/\OmegaHI^{\rm (fid)} = \bHI/\bHI^{\rm (fid)} = \alphaNL = \alphaFoG=1$, to highlight the effect of each parameter on the power spectrum shape. These shapes indicate that the first two parameters will be highly degenerate in a Bayesian analysis, while the other two will also exhibit degeneracies but at a weaker level.
    }
    \label{fig:pk_parameter_dependence}
\end{figure*}

The lower panel of \cref{fig:templates} shows each template normalized by its value at the lowest measured bandpower ($k=0.41\ihMpc$), to highlight the scale-dependence of each term. Terms involving the nonlinear matter power spectrum have larger high-$k$ power than those involving the linear power spectrum, as expected. The different powers of $\mu$ in the ``${\rm hh}$", ``{\rm hv}", and ``{\rm vv}" terms also lead to slightly different shapes. This is because the 2d power spectrum is observed within different ranges of $k_\parallel$ and $k_\perp$ (specifically, $0.37\ihMpc < k_\parallel < 1.58\ihMpc$ and $0.082\ihMpc < k_\perp < 0.40 \ihMpc$). Since $\mu \equiv k_\parallel / k$ with $k \equiv (k_\parallel^2 + k_\perp^2)^{1/2}$, a different range of $\mu$ values contributes to each bin in  $k$.  Changing the power of $\mu$ in the 2d power spectrum therefore has a different effect on the $\mu$-averaged power spectrum at each~$k$.

\cref{fig:pk_parameter_dependence}  displays how the full power spectrum model depends on each of its 4 parameters. The left panels show the model itself, while the right panels show each model curve divided by a ``base" model with $\alpha_\Omega = \alpha_b = \alphaNL = \alphaFoG = 1$. 
The only effect of varying $\OmegaHI$ is to rescale the amplitude. This is also the main effect of varying $\bHI$, although there is a slight scale-dependence because the input cosmological power spectrum is proportional to $[\bHI+f\mu^2]^2$ instead of $\bHI^2$, and \cref{fig:templates} showed that terms with different powers of $\mu$ lead to slightly different shapes in the $\mu$-averaged spectrum. On the other hand, the nonlinear parameters $\alphaNL$ and $\alphaFoG$ have a stronger effect on the shape of the power spectrum.
This implies that we should expect a near-perfect degeneracy between $\OmegaHI$ and $\bHI$, and weaker but non-negligible degeneracies between the other parameters. The results of our Bayesian analysis in \secref{sec:parameter_constraints:posteriors} confirm these expectations.

\section{Simulations}
\label{sec:simulations}

\subsection{Method}
\label{sec:simulations:method}

We generate simulated sky maps and visibilities following a procedure similar to that described in \cite{chimestacking}. We review this procedure here, and include further details in Appendix~\ref{app:sim_details}. Instead of generating \tcm intensity realizations directly from the power spectrum in \cref{eq:P21-theory,eq:Pm-model}, we 
generate realizations of density and velocity-gradient fields whose power spectra can be assembled into the full expression after multiplying by the appropriate prefactors.

We begin with an input matter power spectrum in real space, evaluated at $z=1$.
We transform this power spectrum into real-space correlation functions $\xi(r)$ at $z=1$ for the matter overdensity $\deltam$, a rescaled gravitational potential $\phi$, and their cross-correlation.
We then transform these correlation functions into multi-distance angular power spectra $C_\ell(\chi,\chi')$, evaluated at comoving distances $\chi$, $\chi'$ from an observer, still on a constant-redshift slice at $z=1$.
As part of this transformation, we take line-of-sight derivatives of the $\phi$ leg of the correlation function to convert from $\phi$ to the Kaiser term $\mu^2 \deltam$, and apply Finger-of-God damping by convolving with the position-space version of the kernel from \cref{eq:DFoGk}.

These angular power spectra are integrated over
line-of-sight--distance profiles corresponding to CHIME's $390\,{\rm kHz}$ channels.
We use the results to generate random Gaussian HEALPix maps of $\deltam$ and $\mu^2\deltam$ for each frequency channel. To approximately incorporate evolution along the lightcone, we use our fiducial cosmology's linear growth factor $D^+(z)$ to scale each $\deltam$ and $\mu^2\deltam$ map from $z=1$ to the redshift corresponding to the center of each frequency channel.\footnote{This does not exactly capture time-evolution of the power spectrum at nonlinear scales, but given that our observations cover a narrow redshift range and our treatment of nonlinearities is phenomenological, this growth-factor scaling is sufficient for our purposes in this work.} We then form $\deltaHI$ maps by scaling the $\deltam$ maps by the {\HI} bias $\bHI(z)$, scaling the $\mu^2\deltam$ maps by the logarithmic growth rate $f(z)$, and adding the two.
Finally, we convert the $\deltaHI$ maps to \tcm brightness temperature by scaling by the mean brightness temperature from \cref{eq:Tbarcompact}.
In Appendix~\ref{app:sim_details:validation:flat_sky}, we validate the outputs of this procedure by comparing with an alternative simulation pipeline based on the flat-sky approximation.

To convert simulated maps into visibilities, we use the $m$-mode formalism from \cite{shaw2015}. This formalism relates the Fourier transform of the visibilities in local Earth rotation angle $\phi$,
\beq
V^{\mathrm{m\text{-}mode}}_{pfenm}
	= \int \frac{d\phi}{2\pi} e^{-jm\phi} V_{pfen\phi}\ ,
\eeq
to the spherical harmonic coefficients of a simulated Stokes-$I$ map, $a_{f \ell m}$, via
\beq
V^{\mathrm{m\text{-}mode}}_{pfenm} 
	= \sum_\ell B_{pfen \ell m}\, a_{f \ell m} \ ,
\eeq
where the beam transfer matrix $B_{pfen \ell m}$ incorporates the telescope's primary beam and the interferometric phase associated with different points on the sky.
In the above equations, $p$ is the polarization product (XX or YY) for the baseline of interest, $f$ is frequency channel, and $e$ and $n$ are EW and NS components of the baseline vector in units of cylinder spacing ($\Delta_{\rm EW} = \SI{22}{m}$) and feed spacing ($\Delta_{\rm NS} = \SI{0.3048}{m}$) respectively.
We compute beam transfer matrices for CHIME's feed layout and the \texttt{rev\_03} primary beam model from \citet{chime-auto-paper} using the \texttt{driftscan} package \citep{driftscan}. The natural Earth-rotation-angle resolution $\Delta\phi$ associated with the maximum $m$-mode of the simulated maps is coarser than the time resolution of the re-gridded observed visibilities, so we zero-pad in $m$ before inverse-Fourier transforming $V^{\mathrm{m\text{-}mode}}_{pfenm}$ to obtain $V_{pfen\phi}$.

To convert the simulated visibilities from Kelvin to Janskys, we apply the following conversion:
\beq
\left. V_{pfen\phi} \right|_{\rm Jy}
	= \frac{2 \times 10^{26} \kB \nu_f^2}{c^2} 
	\frac{\Omega_{pf}}{\mathcal{A}_{pf}(\theta_{\rm ref}, 0)}
\left. V_{pfen\phi} \right|_{\rm K}\ ,
\eeq
where $\Omega_{pf}$ is the primary power beam solid angle computed from our beam model and $\mathcal{A}_{pf}(\theta_{\rm ref}, 0)$ is the primary power beam evaluated at a reference declination~$\theta_{\rm ref}$ and zero hour angle. This conversion normalizes the visibilities such that the inferred amplitude of a point source at $\theta=\theta_{\rm ref}$ on meridian is equal to the flux of the source, consistent with CHIME's calibration strategy \citep{CHIMEoverview}.

To obtain a simulated \tcm power spectrum, we process the simulated visibilities using the same steps as for the observed visibilities in \citet{chime-auto-paper}, with the exception of a modified foreground filtering step. For the data, a DAYENU foreground filter \citep{ewall-wice2021} is applied to each of the chosen 94 nights of CHIME data before averaging them together. Instead of simulating full 94-night datasets, we simulate a single set of visibilities, and then apply an effective foreground filter that incorporates the masking, filtering, Earth-rotation-angle-rebinning, and averaging that is applied to the observations (see Section~3 of \citealt{chime-auto-paper} for a description of how this filter is constructed).

To convert pipeline outputs into physical units (${\rm K}^2\, h^{-3}\,{\rm Mpc}^3$), we follow the same procedure as \citet{chime-auto-paper}: generating a simulation with a flat input power spectrum ($P_{21}(k) = 1\,{\rm K}^2\, h^{-3}\,{\rm Mpc}^3$), propagating it through the power spectrum pipeline, and using the result to normalize other pipeline outputs. The same normalization is applied to simulations and data in all figures in this paper.

\subsection{Validation set}
\label{sec:simulations:validation}

In addition to the templates described in \secref{sec:modelling:template}, we generate a set of 20 simulations that we use for validating various aspects of our modelling and interpretation pipeline. The model parameters for these simulations are drawn from a Latin hypercube sampling of the 4-dimensional parameter space ($\OmegaHI$, $\bHI$, $\alphaNL$, and $\alphaFoG$), with bounds $[0.2p_{\rm fid}, 5p_{\rm fid}]$ for each parameter $p$ with fiducial value $p_{\rm fid}$. These ranges were chosen based on conservative priors for where each parameter value is expected to lie.

\begin{figure}[t]
   \centering \includegraphics[width=\linewidth, keepaspectratio, trim = 0 0 0 0]{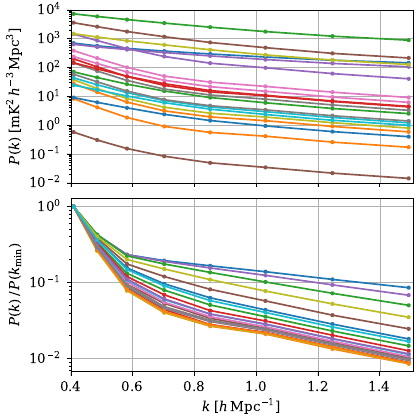}
    \caption{%
    \emph{Upper panel}: \tcm power spectra measured from 20 validation simulations, with model parameters drawn from a Latin hypercube in the 4D parameter space. The power spectra span a wide range of normalizations due to the strong dependence of the normalization on each model parameter.
    \emph{Lower panel}: Each power spectrum is normalized to its value at $k_{\rm min}=0.41\ihMpc$. This demonstrates the wide range of power spectrum shapes spanned by the simulations. These simulations are used for testing
    the robustness of the choices in our Bayesian analysis (\secref{sec:posterior_estimation:validation})
    and the accuracy of our model evaluation pipeline (\secref{sec:parameter_constraints:systematics:template_accuracy}).
    }
    \label{fig:validaton_sim_spectra_bandfull}
\end{figure}

In the upper panel of \cref{fig:validaton_sim_spectra_bandfull}, we show the \tcm power spectra measured from these 20 simulations. Their overall amplitudes span 4 orders of magnitude; this is because each of the 4 model parameters has the effect of rescaling the entire power spectrum (in addition to affecting the shape in different ways -- see \secref{sec:modelling:template}), and therefore the power spectrum normalization is very sensitive to the value of each parameter. When we use these simulations for mock data analyses in \secref{sec:posterior_estimation:validation}, we rescale the noise covariance in each mock analysis so that the S/N is equal to that of the real data, to ensure that prior-volume and projection effects in the mock analyses are similar to those for the data.

In the lower panel of \cref{fig:validaton_sim_spectra_bandfull}, we show the 20 power spectra normalized by their values at the lowest bandpower, $k=0.41\ihMpc$. This demonstrates that the simulations cover a wide range of power spectrum shapes, making them a useful tool for testing the generality of our pipelines.

\subsection{Mock observations of IllustrisTNG simulations}
\label{sec:simulations:tng}

In this work, we use the IllustrisTNG suite of hydrodynamical simulations \citep{TNGa,TNGb,TNGc,TNGd,TNGe} in three ways. 
First, in \secref{sec:posterior_estimation:validation} we verify that our model is flexible enough to provide a good fit to {\HI} power spectra measured from these simulations, to provide evidence that our model spans a realistic range of behavior.
Second, in \secref{sec:parameter_constraints:tngmock} we demonstrate that our model provides unbiased constraints on the power spectrum amplitude if IllustrisTNG power spectra are used as mock data.
Third, in \secref{sec:tng} we compare IllustrisTNG's predictions for the {\HI} power spectrum to the measurements from \citet{chime-auto-paper}.
In each case, we use CHIME simulations generated from IllustrisTNG {\HI} power spectra. We describe the details of how those simulations are generated in Appendix~\ref{app:tng}, but we provide a brief summary here.

We make use of the TNG100-1 and TNG300-1 simulation runs (hereafter referred to as TNG100 and TNG300), which have box side lengths equal to $75\hMpc$ and $205\hMpc$ respectively.\footnote{We do not consider TNG50, which has $L_{\rm box}=35\hMpc$, because the fundamental Fourier wavenumber of this box, $k_{\rm f} = 2\pi/L_{\rm box} = 0.18\ihMpc$, is greater than the minimum wavenumber of our measurements in the transverse direction ($k_{\perp,{\rm min}} = 0.083\ihMpc$). Therefore, this box does not contain the full range of modes that is needed for a comparison with the measurements.}
We post-process the $z=1$ snapshots of each run to evaluate the density of $\HI$ on a regular grid, add redshift-space distortions based on particle velocities along a specific axis, and compute the {\HI} power spectra as a function of $k_\parallel$ and $k_\perp$. We then smooth the measurements and extrapolate to lower wavenumbers that are not contained within each simulation box, to avoid simulation sample variance and missing modes propagating into downstream simulation products. We transform the resulting power spectra into multi-distance angular spectra, integrate over the CHIME frequency channel profile, and feed the results into the CHIME simulation pipeline described in \secref{sec:simulations:method}. Finally, we rescale the results to transform the {\HI} power spectra at $z=1$ into \tcm power spectra at the mean redshift of each band ($z=1.16$ for the full band, and $z=1.08$ and $1.24$ for the upper and lower sub-bands respectively).

\section{Posterior estimation}
\label{sec:posterior_estimation}

\subsection{Method}
\label{sec:posterior_estimation:method}

We estimate the posterior distribution of the model parameters $\vec{\theta}$ governing the \tcm power spectrum using a Bayesian framework. The inference combines the measured power spectrum data vector $\vec{d}$, its associated noise covariance, and the theoretical model described in Section~\ref{sec:modelling}, with the posterior given by
\begin{equation}
P(\vec{\theta} | \vec{d}) = \frac{1}{Z}\, \mathcal{L}(\vec{d} | \vec{\theta}) \, \pi(\vec{\theta})\ ,
\end{equation}
where $\mathcal{L}(\vec{d} | \vec{\theta})$ is the likelihood function, $\pi(\vec{\theta})$ represents the prior distribution of the parameters, and $Z$ is the normalization constant ensuring that the posterior integrates to unity. 

We assume that the noise in the measured power spectra is Gaussian distributed and characterized by the noise covariance matrix $\CNoise$. The likelihood function therefore takes the standard multivariate Gaussian form:
 \begin{equation}
 \mathcal{L}(\vec{d} |\vec{\theta})
 = \frac{1}{|2\pi \CNoise|^{1/2}}
 \exp \left(-\frac{1}{2}\chi^2(\vec{\theta})\right)\ ,
 \end{equation}
 with
 \begin{equation}
 \label{eq:chi2theta}
 \chi^2(\vec{\theta})
 = \left[ \vec{d} - \vec{s}(\vec{\theta}) \right]^{T}
 \CNoise^{-1}
 \left[ \vec{d} - \vec{s}(\vec{\theta}) \right]\ .
 \end{equation}
Here $\vec{d}$ is the data vector of the binned auto-power spectrum measurements, and $\vec{s}(\vec{\theta})$ is the model prediction evaluated at parameter values $\vec{\theta} = [\alpha_{\Omega}, \alpha_{b}, \alphaNL, \alphaFoG]$ (see \cref{eq:OmegaHImodel,eq:bHI-model} for the definitions of $\alpha_\Omega$ and $\alpha_b$). The noise covariance was estimated from 1000 realizations of Gaussian noise consistent with the fast-cadence noise estimates obtained from the CHIME real-time pipeline. 
To account for the bias introduced by the finite number of noise realizations used to estimate $\CNoise$, we multiply $\CNoise^{-1}$ by the “Hartlap factor”, $(n_{\mathrm{mocks}} - n_{\mathrm{data}} - 2)/(n_{\mathrm{mocks}} - 1) \approx 0.99$, when evaluating Equation~\ref{eq:chi2theta} \citep{hartlap2007}.
We explore the posterior distribution using Markov Chain Monte Carlo (MCMC) sampling with the affine-invariant ensemble sampler \texttt{emcee} \citep{foreman-mackey2013}. The analysis employs 12 independent walkers, each evolved for $5 \times 10^6$ steps. The initial 15\% of samples from each walker are discarded as burn-in and the convergence of the chains is assessed using the multivariate Gelman-Rubin diagnostic ($\hat{R}$) \citep{BrooksGelman1998}, ensuring that $\hat{R}-1 < 0.01$ for each chain. We use broad, uniform priors for the parameters: $\alpha_{\Omega} \sim \mathcal{U}(0, 20)$, $\alpha_{b} \sim \mathcal{U}(0,10)$, $\alphaNL\sim \mathcal{U}(0,5)$, and $\alphaFoG \sim \mathcal{U}(0,10)$. The limits on $\alphaNL$ and $\alphaFoG$ are selected to avoid prior volume effects and verified through validation tests using simulated data. Further details and the impact of these prior choices are discussed in Section~\ref{sec:posterior_estimation:validation} and \ref{sec:parameter_constraints:prior_impact}.

A prominent feature of the \tcm power spectrum model in \cref{eq:P21-theory} is the strong degeneracy between the neutral hydrogen density parameter $\OmegaHI$ and the {\HI} bias $\bHI$. As illustrated in Figure~\ref{fig:pk_parameter_dependence}, both parameters primarily act to scale the overall amplitude of the measured power spectrum. The origin of this degeneracy can be seen by considering the linear-theory form of the model,
 \begin{align}
 P_{21}(k, \mu; z)
 &\propto \OmegaHI(z)^2
 \left[ \bHI(z) + f(z)\mu^2 \right]^2
 P_{\mathrm{m}}(k, z)\ ,
 \label{eq:P21-theory_linear}
 \end{align}
which shows that the power spectrum amplitude scales roughly as $\OmegaHI^2 \bHI^2$ to leading order. Consequently, the data are primarily sensitive to the combination $\OmegaHI \bHI$, leading to a pronounced correlation between these parameters in the posterior distribution. Transforming the constraints into the $(\OmegaHI \bHI,\OmegaHI)$ plane largely removes the curved degeneracy, yet a residual linear correlation between $\OmegaHI \bHI$ and $\OmegaHI$ remains. This residual trend arises naturally from the Kaiser redshift-space distortion term $f\mu^2$, which couples the clustering amplitude to the line-of-sight velocity field. For CHIME, whose frequency resolution exceeds its angular resolution and where foreground filtering suppresses modes with small~$k_\parallel$, the effective sensitivity is biased toward modes with large $\mu$. Furthermore, as both $\bHI$ and $f$ are of order unity, the contribution of the Kaiser term is important and cannot be neglected.

Given that the \tcm signal amplitude is governed by the product $\OmegaHI^2(\bHI + f\mu^2)^2$, it is convenient to define a single effective amplitude parameter for the angle-averaged power spectrum,
 \begin{equation}
 \label{eq:ps_amplitude}
 \AHIps \equiv 10^{6}\, \OmegaHI^{2}\left(\bHI + \langle f\mu^{2}\rangle \right)^{2}\ ,
 \end{equation}
 where $\langle f\mu^2\rangle$ represents the sensitivity-weighted angular average of $f\mu^2$. Our notation follows that of \citet{chimestacking}, where $\AHIps$ corresponds to the square of the $\AHI$ parameter used to characterize the amplitude of the CHIME–eBOSS stacking signal in that work.
 This parameter encapsulates the effective amplitude of the power spectrum that is most directly constrained by the data. To be precise, the angle-averaged power spectrum depends on both $\langle f\mu^{2}\rangle$ and $\langle f^{2}\mu^{4}\rangle$ through $P_{21}(k, z) \propto \OmegaHI^{2} 
\left[ 
\bHI^{2} + 2 \bHI \langle f \mu^{2} \rangle 
+ \langle f^{2} \mu^{4} \rangle 
\right]$, which can be seen by expanding the square in \cref{eq:P21-theory_linear} and taking the angular average of each term. 
However, because CHIME’s $\mu$ coverage is restricted to predominantly line-of-sight modes ($\mu \gtrsim 0.6$), the term $\langle f^{2}\mu^{4}\rangle$ can be accurately approximated by $\langle f\mu^{2}\rangle^{2}$, with our analysis showing that the two are consistent to within 2\%, introducing a negligible bias over the accessible range of modes.

To obtain the posterior distribution of $\AHIps$, the value of $\langle f\mu^{2}\rangle$ is estimated directly from the MCMC chains through an eigenanalysis of the posterior covariance between $\OmegaHI \bHI$ and $\OmegaHI$, following the approach used in \citet{chimestacking}. Specifically, the direction of minimum variance in this two-dimensional parameter space defines the linear relation between the two quantities, from which $\langle f\mu^{2}\rangle$ is derived. This optimization is performed using chains in which the non-linear parameters are fixed to their fiducial values, as this setup provides a clearer separation of the degenerate parameters and yields a more robust estimate of $\langle f\mu^{2}\rangle$. The value of $\langle f\mu^{2}\rangle$ derived from the MCMC chains ($0.761$) agrees well with that obtained from the direct angular averaging of the theoretical model ($0.777$), differing by only about $2\%$. This close agreement confirms the validity and internal consistency of the effective treatment adopted in our analysis.

\subsection{Validation}
\label{sec:posterior_estimation:validation}

We have performed a series of validation tests to verify the robustness of the inferred constraints under our fiducial prior choices. These tests were carried out using the set of 20 mock validation simulations described in Section~\ref{sec:simulations:validation}. For each mock realization, synthetic HI power spectra were generated and analyzed through full MCMC inference using the fiducial priors outlined in Section~\ref{sec:posterior_estimation:method}. As discussed in the previous section, the power spectrum model exhibits strong degeneracies among its parameters, implying that the magnitude of the prior volume effect is highly sensitive to the S/N of the measurement. Therefore, when performing MCMC analyses on the validation simulations, we rescaled the noise covariance matrix of each mock by a constant factor, such that the S/N of each mock matched that of the observational data. This rescaling ensures that the validation tests probe the same statistical regime as the real data; otherwise, the mock analyses would exhibit significantly different prior volume effects due to differences in their S/N.

For each validation simulation, we then examined whether the true value of the power spectrum amplitude parameter $\AHIps$ was recovered within the 68\% credible interval of the posterior distribution. Across all 20 simulations, the true values were consistently enclosed within this interval, demonstrating that our fiducial priors yield statistically reliable $\AHIps$ constraints. A more quantitative discussion of how variations in prior assumptions influence the posterior of $\AHIps$ is presented in Section~\ref{sec:parameter_constraints:prior_impact}.

As discussed in \secref{sec:simulations:tng}, we also use IllustrisTNG-derived \tcm auto power spectra, $P^{\rm TNG}_{21, \rm obs}(k)$, to validate our model. 
To ensure that our model validation is subject to prior volume effects at the same level as the data, 
 we scale $P^{\rm TNG}_{21, \rm obs}(k)$ for TNG100 and TNG300 by a factor that ensures that the scaled power spectra, $P^{\rm sTNG}_{21, \rm obs}(k)$, have the same signal to noise ratio as the data. Unlike the 20 validation simulations mentioned earlier in this section, $P^{\rm sTNG}_{21, \rm obs}(k)$ provides a synthetic observation that is not derived from our model and hence provides a qualitatively separate validation of our model.
(In \secref{sec:tng}, we carry out a comparison between the power spectrum measured in \citealt{chime-auto-paper} and $P^{\rm TNG}_{21, \rm obs}(k)$, but that comparison is separate from the validation test we discuss here.)

We validate the posterior distributions for these rescaled IllustrisTNG spectra and our power spectrum model using posterior predictive checks. We first draw 1000 random parameter samples from the posterior distribution and obtain their power spectrum predictions using our model. For each sample, we further obtain 500 data replicates by adding random noise realizations. These replicates are used to obtain standard $\chi^2$ distributions with which we compare the $\chi^2$ between the IllustrisTNG power spectrum and the model predictions corresponding to the respective posterior draws. This results in 1000 $p$-values, one for each posterior sample. We use the mean of these $p$-values as our posterior-predictive check statistic. We find the average $p$-values to be 0.79 and 0.83 for TNG100 and TNG300, respectively. This implies that the power spectrum model is a good description of $P^{\rm sTNG}_{21, \rm obs}(k)$ at the level of current statistical uncertainties. 
\section{Parameter constraints}
\label{sec:parameter_constraints}

\subsection{Posteriors}
\label{sec:parameter_constraints:posteriors}

\begin{figure*}
   \centering \includegraphics[width=0.6\linewidth, keepaspectratio, trim = 0 0 0 0]{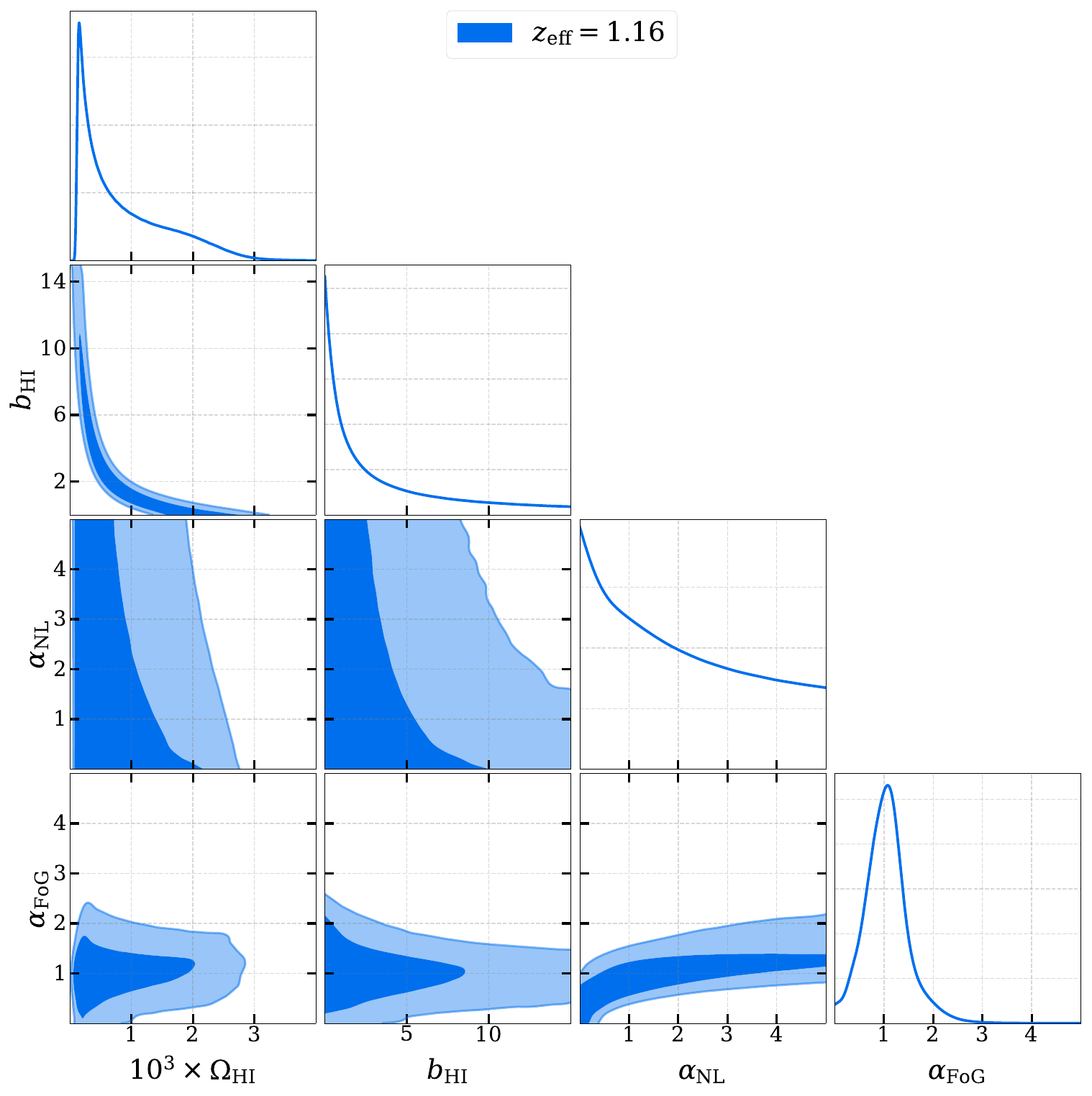}
    \caption{ Posterior distributions of the model parameters obtained from the \tcm auto–power spectrum fit for the full band, covering the frequency range $608.2–707.8$ MHz ($z_{\mathrm{eff}} = 1.16$).  The darker and lighter shaded regions correspond to the 68\% and 95\% credible intervals, respectively.  A strong degeneracy is evident among several parameters, most notably between $\OmegaHI$ and $\bHI$. The nuisance parameter $\alphaFoG$ is well constrained, reflecting the strong scale-dependent effect of varying $\alphaFoG$ over the $k$ range probed.
}
    \label{fig:fullband_posteriors_all}
\end{figure*}

\begin{figure}[t]
   \centering \includegraphics[width=\linewidth, keepaspectratio, trim = 0 0 0 0]{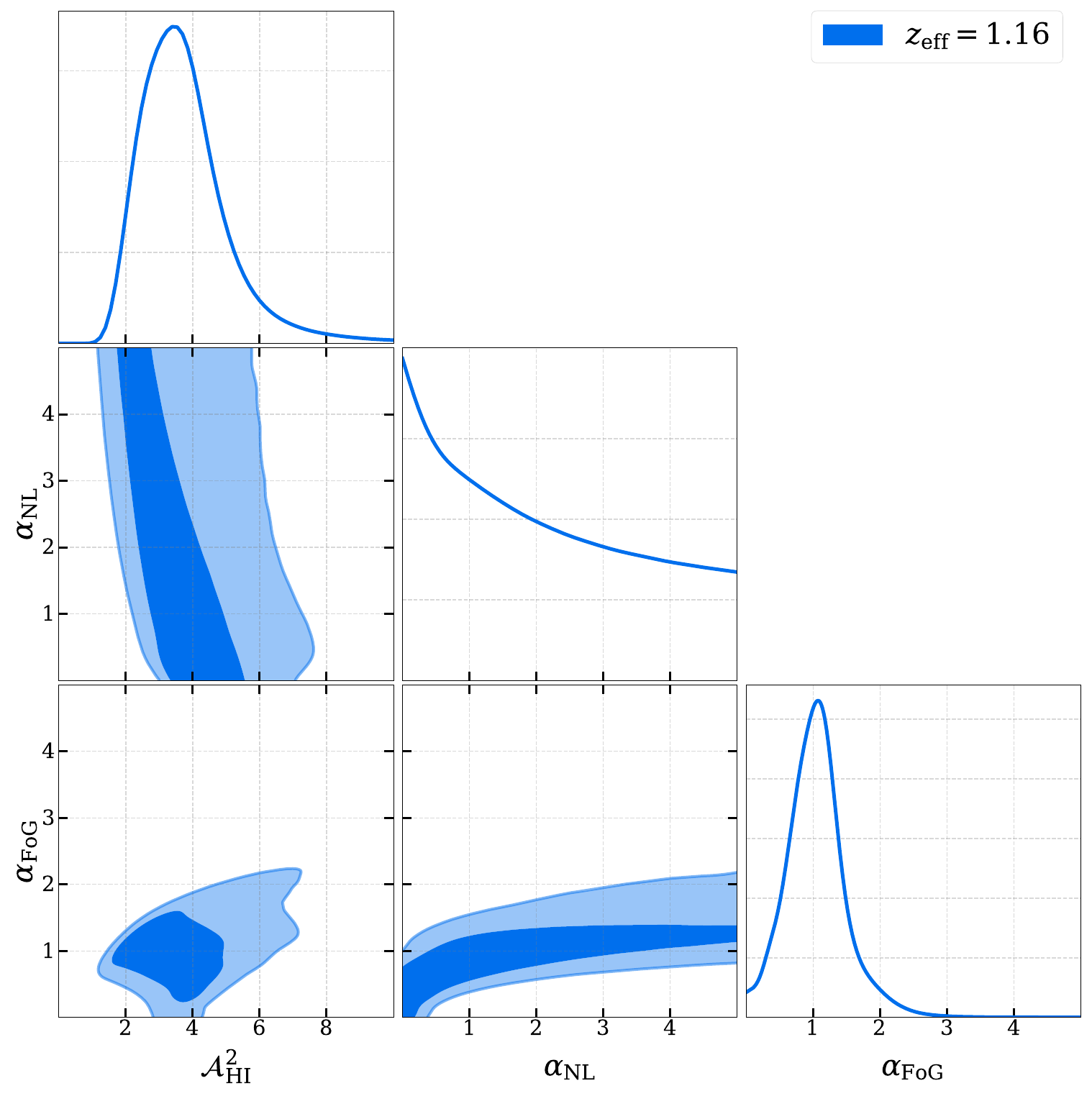}
    \caption{ Same as Figure~\ref{fig:fullband_posteriors_all}, but with the derived power spectrum amplitude parameter $\AHIps$. Transforming to $\AHIps$ yields a cleaner and better-constrained marginalized posterior on $\AHIps$, while revealing a residual degeneracy between $\AHIps$ and $\alphaNL$.
}
    \label{fig:fullband_posteriors_AHI}
\end{figure}

In Figure~\ref{fig:fullband_posteriors_all} we show the marginalized one- and two-dimensional posterior distributions of the model parameters for the full-band fit, corresponding to an effective redshift of $z_{\mathrm{eff}} = 1.16$. The posterior samples were analyzed and visualized with the \texttt{GetDist} package \citep{GetDist}. The two-dimensional contours clearly reveal the strong parameter correlations expected from \cref{eq:P21-theory_linear}, most notably the pronounced degeneracy between $\OmegaHI$ and $\bHI$. As discussed in \secref{sec:posterior_estimation:method}, the data primarily constrain the combination of these parameters through the effective amplitude $\AHIps$, leading to broad marginalized posteriors for the individual quantities. To better capture the amplitude of the \tcm power spectrum, we reparameterize the posterior in terms of the derived quantity $\AHIps$, as defined in Equation~\ref{eq:ps_amplitude}. As described in Section~\ref{sec:posterior_estimation}, we adopt $\langle f\mu^2\rangle = 0.761$ to compute~$\AHIps$,  since it corresponds to the best-constrained definition of this parameter. The resulting posterior distributions, shown in \cref{fig:fullband_posteriors_AHI}, exhibit a substantial reduction in parameter degeneracy, yielding a well-constrained marginalized distribution for~$\AHIps$. A mild residual degeneracy remains with $\alphaNL$, as both parameters modulate the overall amplitude of the power spectrum.

\begin{figure}[t]
   \centering \includegraphics[width=\linewidth, keepaspectratio, trim = 0 0 0 0]{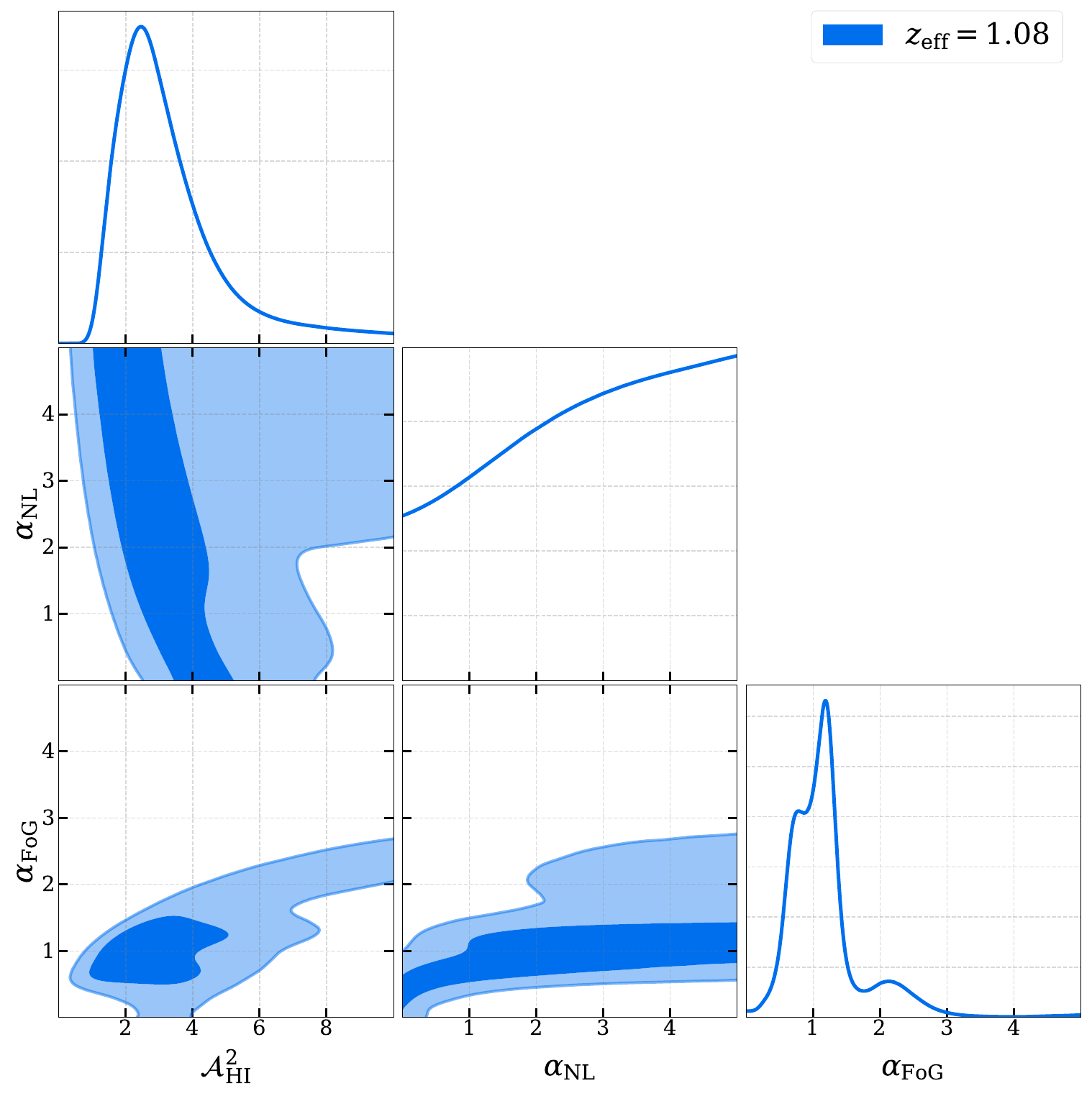}
    \caption{
    Posterior distributions of the power spectrum amplitude parameter $\AHIps$ and the non-linear parameters $\alphaNL$ and $\alphaFoG$ for the upper sub-band, covering the frequency range $658.2 - 707.8$ MHz ($z_{\mathrm{eff}} = 1.08$). The amplitude parameter $\AHIps$ is well constrained and broadly consistent with the full-band results, while mild degeneracies persist with the non-linear parameters. 
    The mild multimodal structure in the marginalized posterior for $\alphaFoG$ is primarily driven by the power spectrum data point at $k\sim 0.85\ihMpc$.
    }
    \label{fig:b1_posteriors_AHI}
\end{figure}

\begin{figure}[t]
   \centering \includegraphics[width=\linewidth, keepaspectratio, trim = 0 0 0 0]{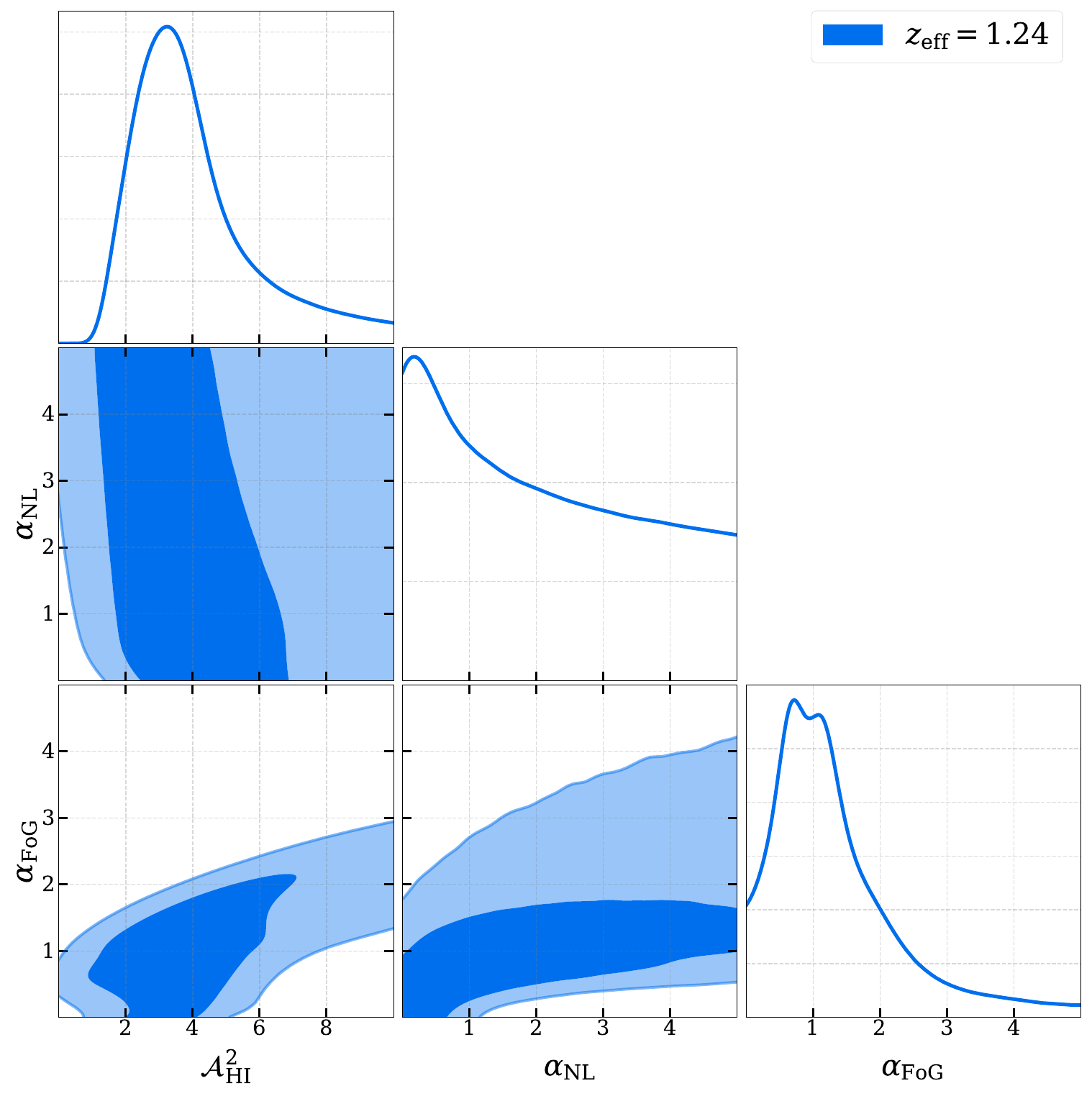}
    \caption{ Posterior distributions of the power spectrum amplitude parameter $\AHIps$ and the non-linear parameters $\alphaNL$ and $\alphaFoG$ for the lower sub-band, covering the frequency range 608.2–658.2 MHz ($z_{\mathrm{eff}} = 1.24$).
    }
    \label{fig:b2_posteriors_AHI}
\end{figure}

In Figures~\ref{fig:b1_posteriors_AHI} and~\ref{fig:b2_posteriors_AHI}, we show parameter posteriors for the two frequency sub-bands analyzed in \citet{chime-auto-paper}, corresponding to effective redshifts $z_{\mathrm{eff}} = 1.08$ and $1.24$, respectively. In both cases, $\AHIps$ remains well constrained, indicating that the amplitude of the \tcm power spectrum is consistently measured across the frequency range. The non-linear parameters $\alphaNL$ and $\alphaFoG$ exhibit broadly similar posterior structures between sub-bands, although the higher-redshift sub-band ($z_{\mathrm{eff}} = 1.24$) shows slightly broader contours. In particular, the posterior of $\alphaFoG$ in the upper sub-band displays a mild multimodal structure. We find that this behavior is primarily driven by the data point at $k \sim 0.85\ihMpc$, whose statistical weight allows multiple FoG damping solutions to provide comparably good fits, thereby introducing secondary maxima in the marginalized $\alphaFoG$ distribution.

The overall quality of the fits, quantified using the $\chi^2$ statistic evaluated at the maximum a posteriori (MAP) point in parameter space, yields values of $1.6$, $6.0$, and $7.1$ for the full band, lower sub-band, and upper sub-band, respectively. 
As described in \citet{chime-auto-paper}, we use a Monte-Carlo based procedure for converting these into $p$-values, finding $0.92$, $0.31$, and $0.25$ for the three bands.
These values indicate statistically acceptable fits in all cases.

The derived MAP estimates of the effective power spectrum amplitude, $\AHIps$, together with their 68\% highest posterior density intervals (HPDIs), are summarized in the third column of Table~\ref{tab:AHIconstraints}. 
The inferred amplitude measurements across the different frequency bands are broadly consistent within their respective statistical uncertainties. Although mild evolution with redshift is expected between the sub-bands, the current S/N is insufficient to detect this evolution conclusively. The posteriors are non-Gaussian and exhibit significant skewness; however, despite the broad credible intervals, the probability that $\AHIps \le 0$ is negligible. Although the non-linear parameters are partially degenerate with the overall amplitude, a non-zero $\AHIps$ is required for any measurable signal. Due to parameter degeneracies, the marginalized constraints on $\AHIps$ are susceptible to prior-volume effects. We discuss this further in \secref{sec:parameter_constraints:prior_impact}.

Furthermore, \citet{chime-auto-paper} compared our inferred amplitude parameter $\AHIps$ from the full-band analysis with the effective {\HI} clustering amplitude $\AHI$ obtained from the CHIME–eBOSS stacking analysis of the same dataset \citep{chimestacking}. The stacking measurement, based on cross-correlation with 33,119 eBOSS quasars in the redshift range $1.01 < z < 1.34$ ($608$–$708$ MHz), yields $\AHI = 1.93^{+2.50}_{-0.93}$ when the non-linear parameters are free and $\AHI = 1.63^{+0.19}_{-0.18}$ when they are fixed to their fiducial values. Reprocessing the power spectrum chains to compute $\AHI$ instead of $\AHIps$, to enable a direct comparison,  \citet{chime-auto-paper} found good agreement between the auto-correlation and cross-correlation results, with consistency at the $1\sigma$ level. This consistency between two independent analyses strengthens the interpretation that the detected \tcm signal is cosmological in origin, rather than being driven by residual systematics, which would primarily affect the auto-correlation measurement. Further details about this comparison are provided in Section 7.10 of \citet{chime-auto-paper}.

\subsection{Systematic uncertainties}
\label{sec:parameter_constraints:systematics}

\begin{deluxetable*}{c c CCCCCCCC}[tb]
    \tablecaption{Constraints on power spectrum amplitude parameter $\AHIps$.\label{tab:AHIconstraints}}
    \tablecolumns{9}
    \tablehead{
    	\multicolumn{2}{c}{Band} &
	\multicolumn{1}{c}{} &
        \multicolumn{5}{c}{Systematic uncertainty} &
        \multicolumn{1}{c}{} 
        \\
        \cmidrule(lr){1-2}
        \cmidrule(lr){4-8}
        \colhead{$\nu$ (MHz)} &
        \colhead{$z$} &
        \colhead{\shortstack[c]{MAP value\\[1pt] $\pm$ statistical\\[1pt] uncertainty}} &
	\colhead{\shortstack[c]{Prediction\\[3pt] accuracy}} &
        \colhead{\shortstack[c]{Template\\[-1pt] sample\\[-1pt] variance}} &
        \colhead{\shortstack[c]{Beam\\[1pt] model}} &
        \colhead{\shortstack[c]{Flux\\[1pt] scale}} &
        \colhead{Total} &
        \colhead{\shortstack[c]{Result with\\[1pt] combined\\[1pt] uncertainty}}
    }
    \startdata
    $608$-$708$ &
    	$1.34$-$1.01$ &
    3.55^{+0.96}_{-1.32} &
	1.3\% &
	1.4\%  &
	15\% &
	8\% &
	\pm 0.61 &
	3.55^{+1.14}_{-1.45} \\
   $658$-$708$ &
    	$1.16$-$1.01$ &
    2.47^{+1.59}_{-1.08} &
	2.2\% &
	1.6\%  &
	15\% &
	8\% &
	\pm 0.43 &
	2.47^{+1.65}_{-1.16} \\
    $608$-$658$ &
    	$1.34$-$1.16$ &
    3.24^{+6.79}_{-2.11} &
	1.7\% &
	1.4\%  &
	15\% &
	8\% &
	\pm 0.56 &
	3.24^{+6.81}_{-2.18} \\
    \enddata
    \tablecomments{See \secref{sec:parameter_constraints:posteriors} for discussion of how the maximum a posteriori (MAP) value and statistical uncertainty are determined, and
    \secref{sec:parameter_constraints:systematics} for descriptions of how each systematic uncertainty contribution is estimated. For the total systematic uncertainty, we add each contribution in quadrature, multiplying percentage values by the MAP value for $\AHIps$. The final column lists the MAP value along with the quadrature sum of statistical and systematic uncertainties.
}
\end{deluxetable*}

The parameter posteriors presented in \secref{sec:parameter_constraints:posteriors} do not incorporate potential sources of systematic error in our analysis. In this subsection, we attempt to quantify the largest systematic uncertainties, and propagate them into additional uncertainty on the constraints on~$\AHIps$. We summarize the results in \cref{tab:AHIconstraints}. We conservatively take each systematic uncertainty to be Gaussian with standard deviation described below, and add each standard deviation in quadrature to obtain a total systematic uncertainty. For uncertainties estimated as a fractional uncertainty on the power spectrum amplitude, we multiply this fractional uncertainty by the MAP value for $\AHIps$ from \secref{sec:parameter_constraints:posteriors} when computing the total uncertainty.

\subsubsection{Accuracy of template-based predictions}
\label{sec:parameter_constraints:systematics:template_accuracy}

The template-based framework described in \secref{sec:modelling:template} may not exactly reproduce the desired predictions, due to numerical inaccuracies in various stages of the simulation or model-evaluation pipelines.
To quantify this, we evaluate the power spectrum prediction at points in parameter space corresponding to each of the 20 validation simulations from \secref{sec:simulations:validation}, and compare the results to the power spectrum measured from each simulation. Across all 20 simulations and all measured bandpowers, 
the root-mean-square fractional difference between the predictions and simulations is $1.3\%$ for the full band, $2.2\%$ for the upper sub-band, and $1.7\%$ for the lower sub-band. These values are within a factor of two of the root-mean-square difference between our reference simulation pipeline and a simpler flat-sky approach (see Appendix~\ref{app:sim_details:validation:flat_sky}), indicating that the discrepancy is consistent with arising from our simulation pipeline rather than the intrinsic accuracy of our template-based prediction framework.
We expect that further refinements of our numerical pipelines would reduce these errors, but since they are subdominant to the statistical uncertainties for all 3 bands, they are at an acceptable level for this work.

\subsubsection{Template sample variance}
\label{sec:parameter_constraints:systematics:template_sample_variance}

Each power spectrum template is computed from a single large-scale structure realization, and will therefore be subject to sample variance based on the finite number of modes in the simulation. To quantify this, we generate 20 realizations of a template with $(\OmegaHI, \bHI, \alphaNL, \alphaFoG)=(\OmegaHI^{\rm (fid)}, \bHI^{\rm (fid)}, 1, 1)$, and compute the ratio of the standard deviation and mean over the 20 realizations at each bandpower. The maximum of this ratio over all bandpowers is $1.4\%$, $1.6\%$, and $1.4\%$ for the full band, upper sub-band, and lower sub-band respectively.

\subsubsection{Beam model}
\label{sec:parameter_constraints:systematics:beam}

Our power spectrum templates assume the (\texttt{rev\_03}) primary beam model from \citet{chime-auto-paper}. This model captures features of the beam at hour angles less than roughly $2^\circ$, while lacking east-west sidelobes at larger hour angles. To assess the impact of this missing information, we compare with an alternative beam model that extends across the entire sky visible to CHIME.

This full-sky beam model is based primarily on solar beam measurements interpolated onto a grid of telescope-frame coordinates $(x,y)$ \citep{solar-beam-paper}. Because the Sun samples only a limited range of declinations, these measurements provide incomplete coverage in $y$. To extend the model over the full $(x,y)$ domain, we perform a singular value decomposition of the measured beam, retain the top three modes, preserve their measured $x$-dependence, and refit their $y$-dependence with cubic B-splines subject to a smoothness regularization. The fit is constrained by both the solar measurements and visibilities averaged over sidereal nights. The visibility fit uses a point-source sky model and is restricted to intercylinder baselines and the RA range considered in this analysis.

For each beam model, we generate simulated \tcm visibilities based on the model parameters used in \secref{sec:parameter_constraints:systematics:template_sample_variance}, and form ringmaps as described in \citet{chime-auto-paper}. The full-sky beam model is only available at 28 sparsely-spaced frequencies within the relevant band (\SIrange{608.2}{707.8}{\mega\hertz}), so we do not apply a foreground filter to these simulations, because our chosen filter cannot easily handle such a sparse sampling. We then compute the variance of the ringmap pixel values at each frequency. For the two simulations, these variances agree to within 15\% at all frequencies, so we assume a 15\% systematic uncertainty on the power spectrum amplitude associated with uncertainties in beam modelling.

\subsubsection{Flux scale}

As described in \citet{CHIMEoverview} and \citet{chimestacking}, CHIME's gain calibration strategy is primarily based on daily observations of Cygnus A, which are compared to a reference model from \citet{perley2017} that interpolates between previous measurements from the Karl G.\ Jansky Very Large Array within the CHIME band. The associated absolute flux scale is estimated to have an uncertainty of $3-5\%$ based on measurements by \citet{baars1977}. This was incorporated into the eBOSS stacking analysis in \citet{chimestacking} by assuming a 4\% systematic uncertainty on the amplitude of the stacking signal. Since the \tcm auto power spectrum is quadratic in the CHIME data, we assume that the flux scale uncertainty adds an 8\% systematic uncertainty to the power spectrum amplitude.

\subsection{Impact of priors}
\label{sec:parameter_constraints:prior_impact}

When performing Bayesian inference in high-dimensional parameter spaces, the choice of prior distributions on nuisance parameters can substantially influence the marginalized posteriors of parameters of interest. Related phenomena are commonly referred to as prior volume or projection effects,
and appear in various contexts in cosmology, including large-scale structure perturbation theory \citep{simon2023-eftpriors,carrilho2023-priors,holm2023-eftpriors,chudaykin2024-eftpriors}, early dark energy \citep{ivanov2020-edepriors,smith2021-edepriors}, weak lensing and galaxy clustering analysis \citep{DES:2026zjp}, and density split statistics \citep{ITEM_prior}.
 In our analysis, the power spectrum model exhibits strong degeneracies among its parameters (see Section~\ref{sec:parameter_constraints:posteriors}), and we are in a relatively low-S/N regime. These conditions enhance the susceptibility of the inferred posteriors to prior volume effects. Consequently, the assumed prior ranges on the nuisance parameters $\alphaNL$ and $\alphaFoG$ can indirectly bias the marginalized constraints on $\AHIps$, potentially leading to systematic offsets in cosmological inference.

\begin{figure}[t]
   \centering \includegraphics[width=\linewidth, keepaspectratio, trim = 0 0 0 0]{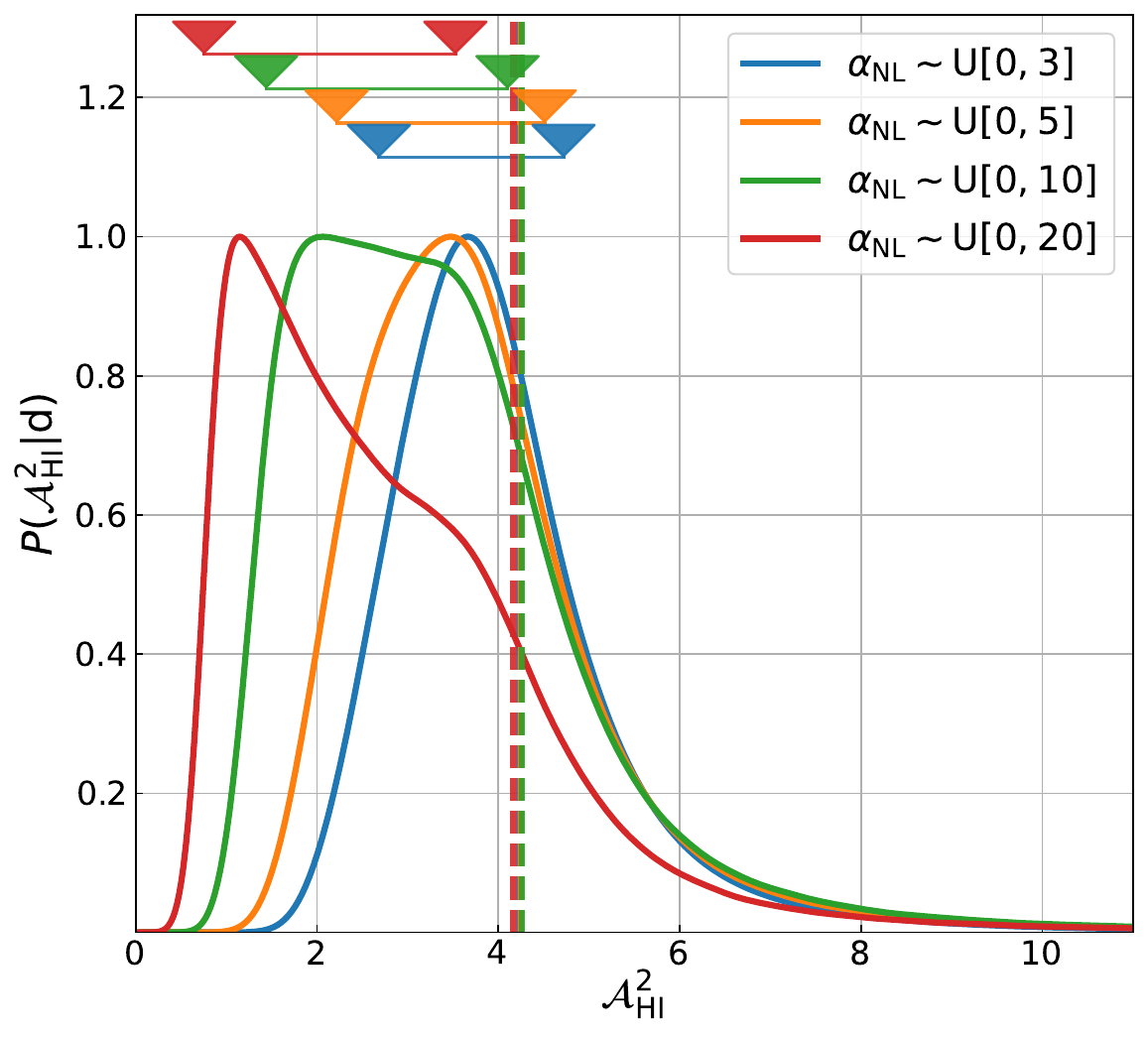}
    \caption{ Effect of the prior range of $\alphaNL$ on the  marginalized constraints of the amplitude parameter $\AHIps$. The vertical dashed lines indicate the best-fit values for each case, and the arrows at the top mark the 68\% highest posterior density intervals (HPDIs). Although the posterior distributions shift systematically with changes in the prior of $\alphaNL$,
    the best-fit values remain consistent across all cases, indicating that the posterior shifts are 
     due to prior volume effects. Based on other validation tests described in the main text, we adopt $\alphaNL \sim \mathcal{U}[0,5]$ in our analysis.
     }
    \label{fig:prior_volume_NL}
\end{figure}

\begin{figure}[t]
   \centering \includegraphics[width=\linewidth, keepaspectratio, trim = 0 0 0 0]{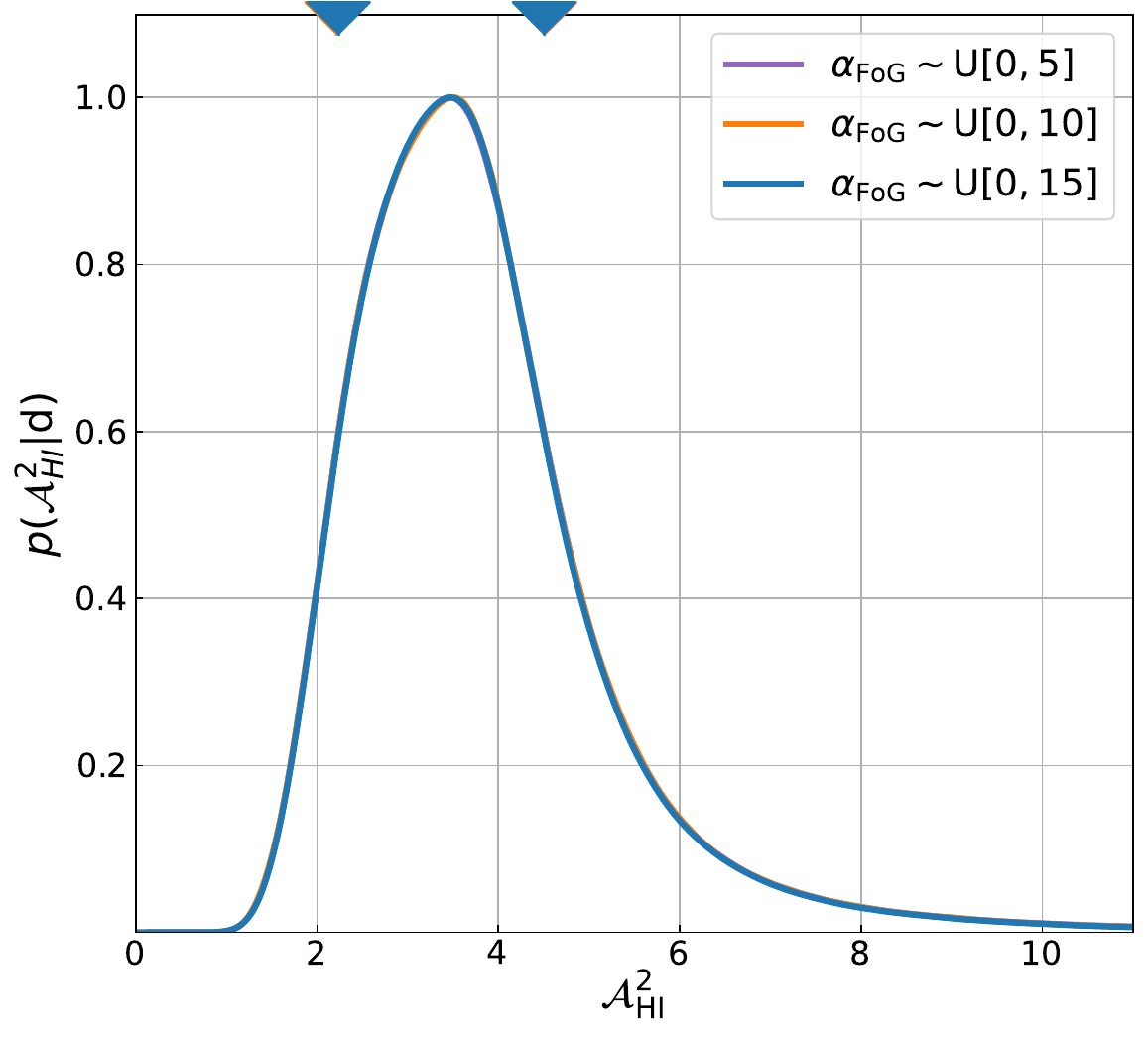}
    \caption{ Same as Figure~\ref{fig:prior_volume_NL}, but for $\alphaFoG$. The posterior distribution of $\AHIps$ remains consistent across all prior choices of $\alphaFoG$, indicating no significant prior volume effect.
    }
    \label{fig:prior_volume_FoG}
\end{figure}

To quantify the impact of prior assumptions, we perform a sensitivity analysis by varying the prior ranges of $\alphaNL$ and $\alphaFoG$ while keeping all other analysis settings fixed, as described in \secref{sec:posterior_estimation}. Figures~\ref{fig:prior_volume_NL} and~\ref{fig:prior_volume_FoG} depict the resulting marginalized posterior distributions of $\AHIps$ under different prior choices. In \cref{fig:prior_volume_NL}, the prior on $\alphaNL$ is varied across $\mathcal{U}[0,20]$, $\mathcal{U}[0,10]$, $\mathcal{U}[0,5]$, and $\mathcal{U}[0,3]$. As the prior volume on $\alphaNL$ contracts, the marginalized posterior for $\AHIps$ shifts systematically toward higher values. The 68\% HPDIs, marked by the colored arrows at the top of the figure, also shift across different priors, demonstrating that prior choices quantitatively affect our amplitude constraints. This behavior is characteristic of prior volume effects in degenerate parameter spaces. The degree of posterior shift provides a direct measure of the sensitivity of our constraints to prior specification and highlights the necessity of establishing well-justified, stable priors on nuisance parameters.

We do so by applying the validation procedure described in \secref{sec:simulations:validation}. Using the $20$ mock realizations, we examine different upper bounds on the uniform prior for $\alphaNL$: $\mathcal{U}[0,40]$, $\mathcal{U}[0,10]$, $\mathcal{U}[0,5]$, $\mathcal{U}[0,3]$ and $\mathcal{U}[0,2]$. For each prior configuration and validation simulation, we assess the frequency with which the true value of $\AHIps$ falls within the 68\% HPDI derived from the posterior. This provides a direct, empirical measure of whether a given prior choice yields statistically reliable constraints. We find that for the prior choices $\mathcal{U}[0,5]$ and $\mathcal{U}[0,10]$, the true value of $\AHIps$ is recovered within the 68\% HPDI for all validation simulations, whereas broader prior ranges fail to satisfy this coverage criterion. Adopting a more stringent requirement based on the 50\% HPDI, only the $\mathcal{U}[0,5]$ prior maintains consistent coverage, while all wider prior choices underperform.

Furthermore, as illustrated in Figure~\ref{fig:prior_volume_NL}, the marginalized posteriors of $\AHIps$ shift systematically as the prior volume of $\alphaNL$ changes, yet the best-fit values remain stable across all tested cases. This behavior confirms that the likelihood itself favors the same region of parameter space, and that the apparent shifts in the marginalized posteriors arise primarily from prior weighting rather than changes in the underlying likelihood structure. Based on these results, we adopt $\alphaNL \sim \mathcal{U}[0,5]$ as the fiducial prior range. The selected prior is sufficiently broad to encompass plausible non-linear effects, yet restrictive enough to prevent over-constraining, unlike narrower priors with upper bounds below $5$. It also enables adequate exploration of the $\AHIps$–$\alphaNL$ degeneracy while remaining consistent with theoretical expectations that very large values of  $\alphaNL$ correspond to an unphysical regime. 

We also vary the prior range of $\alphaFoG$, but find that the resulting posteriors for $\AHIps$ show negligible changes (see \cref{fig:prior_volume_FoG}), indicating that $\alphaFoG$ does not contribute significantly to prior-volume effects in our analysis. We further investigate the impact of prior choices on the remaining parameters, $\alpha_{\Omega}$ and $\alpha_{b}$, and find that the adopted priors for these parameters are sufficiently broad so as not to restrict the degeneracy direction between them. Among these, $\alpha_{\Omega}$ is relatively well constrained, exhibiting no extended tails in its posterior distribution (see \cref{fig:fullband_posteriors_all}). For $\alpha_{b}$, an increased upper bound from $10$ to $40$ has negligible influence on the inferred amplitude parameter $\AHIps$.
Based on these results, we adopt $\alpha_{\Omega} \sim \mathcal{U}[0,20]$, $\alpha_{b} \sim \mathcal{U}[0,10]$, $\alphaNL \sim \mathcal{U}[0,5]$, and $\alphaFoG \sim \mathcal{U}[0,10]$ as the fiducial priors for our analysis. The same prior choices are applied consistently to the two sub-band analyses at $z_{\mathrm{eff}} = 1.08$ and $z_{\mathrm{eff}} = 1.24$.

\subsection{Parameter inference with IllustrisTNG mock}
\label{sec:parameter_constraints:tngmock}

In \secref{sec:posterior_estimation:validation}, we confirmed the flexibility of our power spectrum model by checking that it can provide a good fit to scaled \tcm power spectra derived from the IllustrisTNG simulations, $P^{\rm sTNG}_{21, \rm obs}(k)$. In that test, we rescaled the spectra from the simulations so that they each had the same S/N as the observed power spectrum, to ensure that prior-volume and projection effects are similar in the data and mock analyses. In this subsection, we assess whether mock analyses of these scaled TNG100 and TNG300 power spectra return unbiased constraints on the amplitude parameter $\AHIps$, by comparing the marginalized posterior for this parameter with the known value for each simulation.

These known values are based on measurements of $\OmegaHI$ and $\bHI$ from each simulation. We compute $\OmegaHI(z=1)$ based on the total {\HI} mass of each $z=1$ simulation snapshot, yielding $6.35\times 10^{-4}$ for TNG100 and $5.97\times 10^{-4}$ for TNG300, and rescaling these values from $z=1$ to $z=1.16$ using the redshift-dependence of the fitting function from \cite{crighton2015}, given in \cref{eq:OmegaHIfid}. This yields $\OmegaHI(z=1.16)=6.65\times 10^{-4}$ for TNG100 and $6.25\times 10^{-4}$ for TNG300. For the {\HI} bias in TNG100, we evaluate the fiducial model in \cref{eq:bHI-fitfunc} (which is based on measurements from TNG100 in \citealt{villaescusa-navarro2018}) at $z=1.16$, which gives a value of $1.56$. For TNG300, we evaluate the {\HI} bias fitting function from \citet{foreman2024-HIstoch} (which was derived from field-level perturbation theory fits to TNG300) at $z=1.16$, finding $1.39$. For each simulation, we use these values to compute $\AHIps$, and then scale this number by the S/N scaling factor $f_{\rm S/N}$ that has been applied to the power spectra. We summarize these different values in \cref{tab:TNG_specifications}.

\begin{deluxetable}{c c c c c c}[tb]
    \tablecaption{Parameter values for fits to scaled IllustrisTNG power spectra.\label{tab:TNG_specifications}}
    \tablecolumns{5}
    \tablehead{
        \colhead{Simulation} &
        \colhead{$10^3 \OmegaHI$} &
        \colhead{$\bHI$}&
        \colhead{$f_{\rm S/N}$} &
        \colhead{$\AHIps$}&
        \colhead{Best-fit $\AHIps$}
    }
    \startdata
    TNG100 & $0.665$ & 1.56 & 2.44 & 5.83 & 5.37 \\
    TNG300 & $0.626$ & 1.39 & 4.01 & 7.26 & 5.84 \\
    \enddata
   \tablecomments{The $\OmegaHI$ and $\bHI$ values are based on measurements from $z=1$ simulation snapshots, scaled to $z = 1.16$.
   We rescale IllustrisTNG \tcm power spectra by $f_{\rm S/N}$ so that they have the same S/N as the CHIME measurement, assuming the noise covariance computed from CHIME simulations. The fiducial values of $\AHIps$ are computed assuming $\fmu=0.761$, using $\AHIps = f_{\rm S/N} 10^6 \OmegaHI^2(\bHI^2+\fmu)^2$. The best-fit $\AHIps$ values correspond to the best-fit point in the four-parameter power spectrum model space.
   }
\end{deluxetable}

\begin{figure}[t]
   \centering
   \includegraphics[width=\linewidth, keepaspectratio, trim = 0 0 0 0]{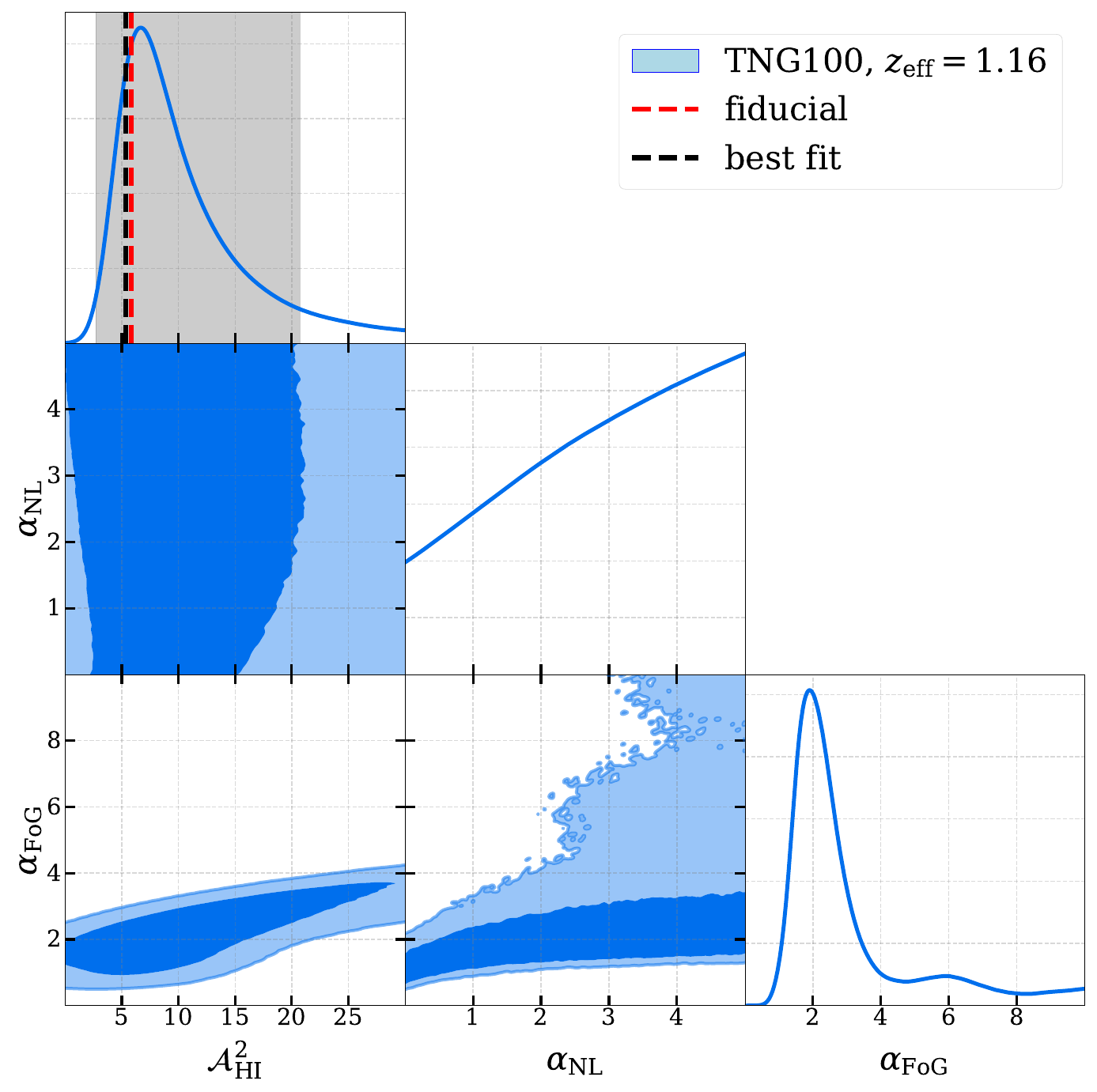}   
   \includegraphics[width=\linewidth, keepaspectratio, trim = 0 0 0 0]{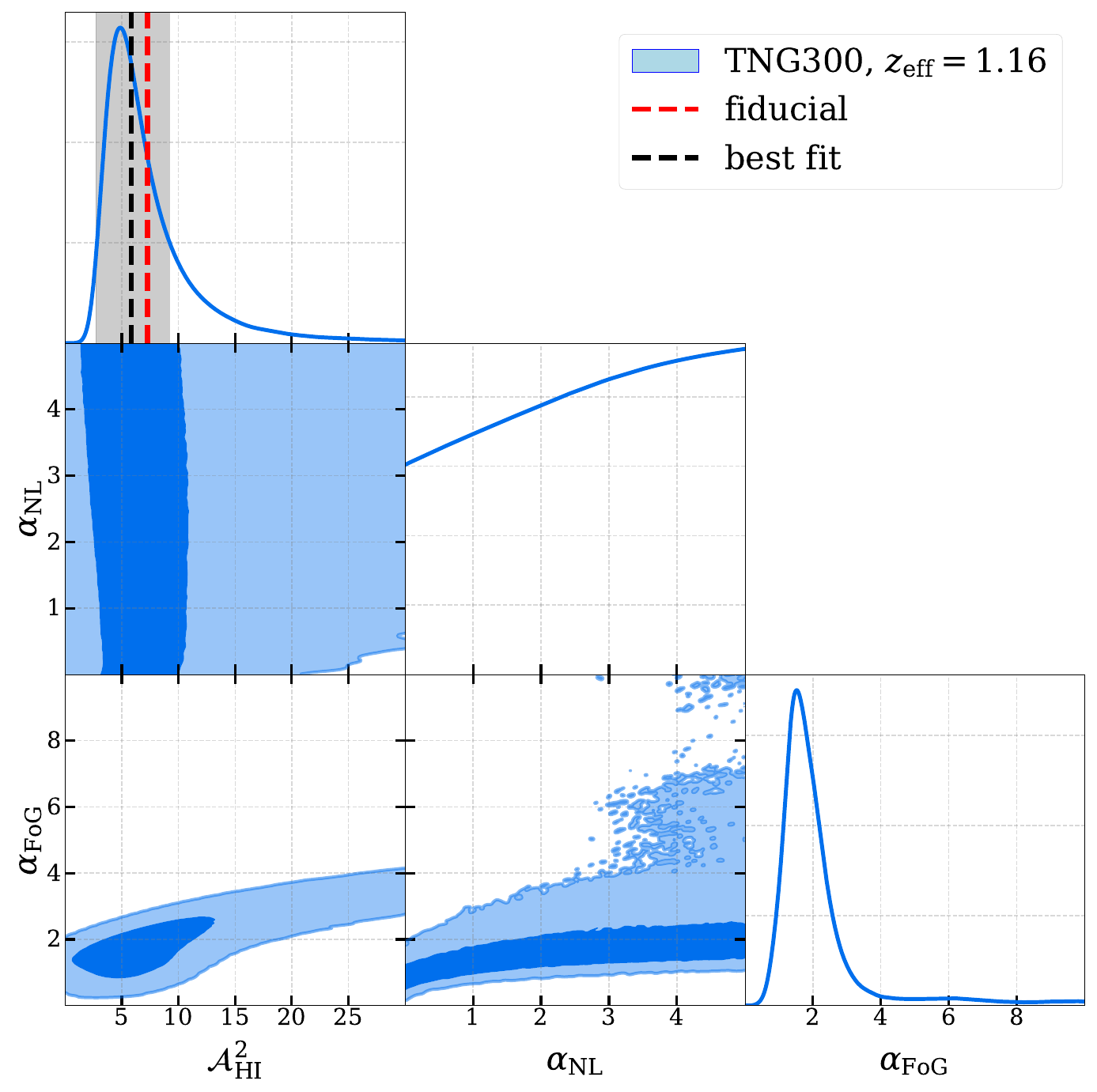}
    \caption{%
    Posteriors distributions of $\AHIps$, $\alphaNL$, and $\alphaFoG$ for mock data derived from the TNG100 (\textit{upper panel}) and TNG300 (\textit{lower panel}) simulations. Red dashed lines denote the known values of $\AHIps$ for each simulation, and black dashed lines denote the best-fit values.  For both TNG100 and TNG300 the fiducial $\AHIps$ value falls within 68\% HPDI of the marginalized distribution indicated by gray shaded region.
   }
    \label{fig:data_tng_params}
\end{figure}

We carry out MCMC analyses of each scaled IllustrisTNG power spectrum $P^{\rm sTNG}_{21, \rm obs}(k)$.
To avoid any random fluctuations, we do not add any noise to these power spectra; however, we do use the same covariance matrix and set of priors as used in our baseline analysis. In \cref{fig:data_tng_params}, we show the marginalized 2D and 1D distributions of $\AHIps$, $\alphaNL$, and $\alphaFoG$ inferred from each scaled power spectrum. Between TNG100 and TNG300, we find the best-fit value of $\AHIps$ is at a similar level of agreement with the fiducial value for each simulation run. The marginalized distributions of $\AHIps$ for TNG100 and TNG300 are in agreement with their respective fiducial values, which fall within the 68\% HPDI region of marginalised $\AHIps$ distribution. Posterior distributions for both TNG100 and TNG300 show a similar type of degeneracy between parameters as seen in the posterior distribution for the data. The 1D marginalised distributions of $\AHIps$ and $\alphaFoG$ are well constrained, and that of $\alphaNL$ is constrained mainly by the prior range. This analysis shows that our four-parameter model is sufficient to return unbiased constraints on the HI power spectrum amplitude when tested with hydrodynamical simulations, at the S/N of the CHIME measurement.

\subsection{Separate constraints on $\OmegaHI$ and $\bHI$}

$\OmegaHI$ and $\bHI$ are crucial ingredients to understand the distribution of HI and can, in principle, be inferred from HI intensity mapping observations. With the power spectrum measurement from \cite{chime-auto-paper}, we can infer the joint distribution of $\OmegaHI$ and $\bHI$; however, the signal to noise ratio and range of scales of this measurement imply that the measurement on its own is unable to break the degeneracy between these two parameters. With external information on either of these two parameters, however, we can obtain an independent constraint on the other parameter. 

\subsubsection{$\OmegaHI$}

\begin{figure}
   \centering \includegraphics[width=\linewidth, keepaspectratio, trim = 0 0 0 0]
   {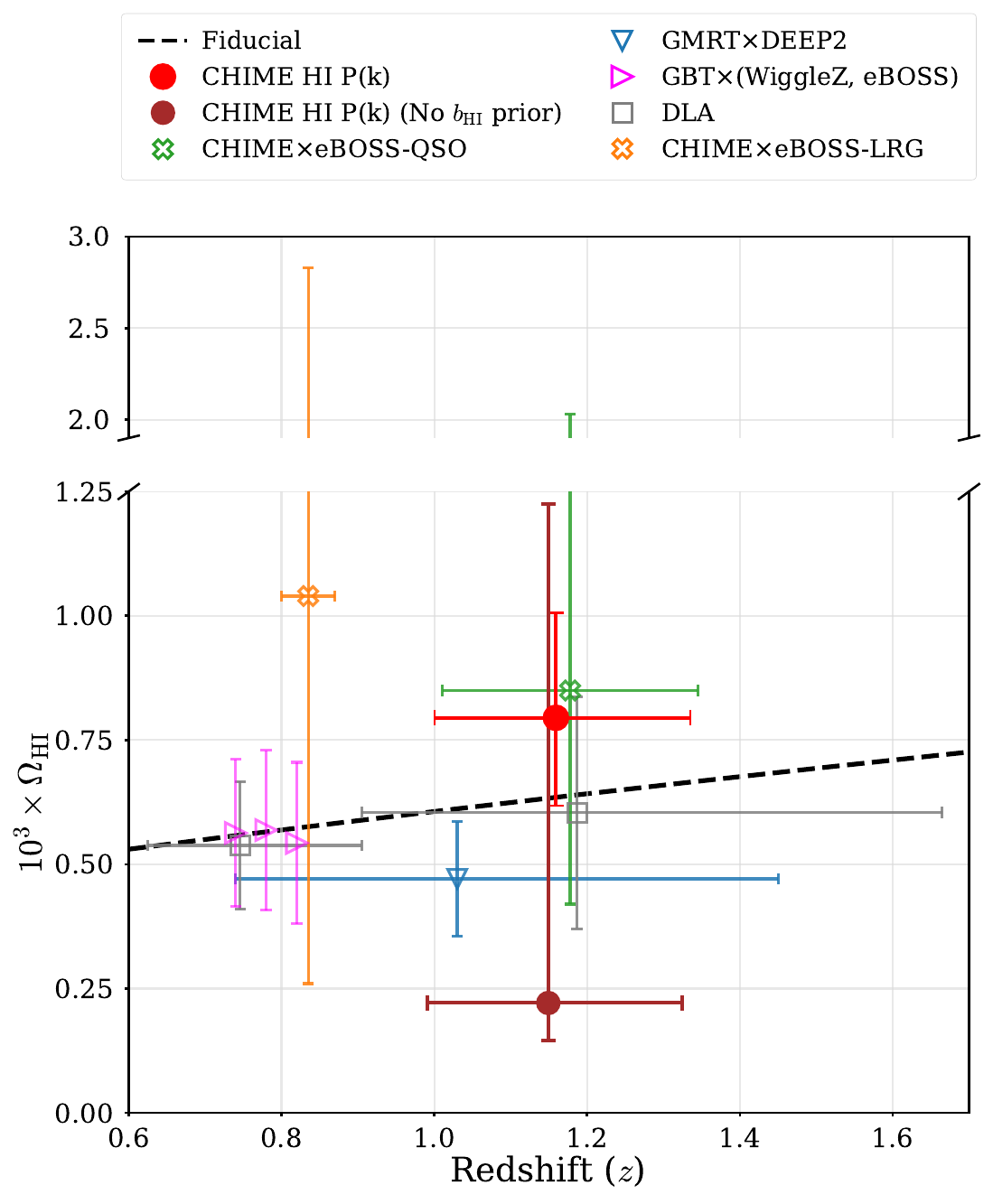}
    \caption{Comparison of the $\OmegaHI$ constraints with (red point) and without (brown point) an additional prior on $b_{\rm HI}$ from this analysis with a representative sample of other $\Omega_{\rm HI}$ measurements in the literature (see main text for descriptions). The green point shows the $\Omega_{\rm HI}$ constraint obtained by stacking the eBOSS-QSO catalog on the same CHIME data used in this analysis. The orange point shows the $\Omega_{\rm HI}$ from stacking the eBOSS-LRG catalog on CHIME data reported in \citet{chimestacking}. The black dashed curve shows the fitting function for $\Omega_{\rm HI}(z)$ from \citet{crighton2015}. We find reasonable consistency between our measurements and those from the literature.
    }
    \label{fig:omega_HI}
\end{figure}

For constraints on $\OmegaHI$, we follow the procedure used in \cite{chimestacking}, where they use a simulation-informed prior on $\bHI$. We use a Gaussian prior centered at a value of $\bHI(z = 1.16)=1.56$ determined from our fiducial model, with standard deviation, $\sigma_{\bHI}$, that is 20\% of the fiducial value, $\sigma_{\bHI}/\bHI(z = 1.16) = 0.2$. We re-weight the posterior samples using the additional prior on $\bHI$ and marginalize over other parameters to obtain constraints on $\OmegaHI$, finding $\OmegaHIMargConstraint$.
In \cref{fig:omega_HI}, we show these constraints with a horizontal errorbar spanning the redshift range of our baseline analysis, $z = 1.01 - 1.34$. We also show the constraint without the additional prior on $\bHI$, $\OmegaHIMargConstraintNoPrior$, to indicate the effect of the additional prior and consistency of the constraints. \citet{chime-auto-paper} also reports the stacking measurement of the eBOSS-QSO catalog on the CHIME data used in the power spectrum analysis. Using this stacking measurement and the same prior on $\bHI$, we obtain $\OmegaHIMargConstraintQSOStack$, consistent with the power spectrum result.

For comparison, we show other measurements of $\OmegaHI$ from the literature in the range $z \approx 0.6 - 1.6$. The CHIMExeBOSS-LRG data point from the stacking analysis in \citet{chimestacking} uses the Gaussian prior on $\bHI$ imposed in that work, with mean $1.30$ and standard deviation equal to 20\% of the mean.
For the GMRT stacking measurement on DEEP2 galaxies from \citet{chowdhury2020} and damped Lyman-$\alpha$ (DLA) measurements from \citet{rao2017}, we convert their $\OmegaHI$ constraints to use the Planck 2018 cosmology used in this work, following the procedure from \citet{chimestacking}.
For the \tcm cross-correlations with WiggleZ and eBOSS galaxies from \citet{wolz2021}, we convert their $\OmegaHI \bHI r$ constraints into $\OmegaHI$ using $\bHI=1.38$ (the value at $z=0.78$ in our fiducial model) and the $r$ values assumed in that work ($r_{\sHI,{\rm Wig}}=0.9$, $r_{\sHI,{\rm ELG}}=0.7$, and $r_{\sHI,{\rm LRG}}=0.6$).

We see reasonable agreement between our $\OmegaHI$ constraint and these other estimates. Further, our constraint is consistent with the fitting function from \citet{crighton2015} (Equation~\ref{eq:OmegaHIfid}), which is based on a compilation of measurements from direct {\HI} galaxy surveys, stacking, and damped Lyman-$\alpha$ catalogs.

\subsubsection{$\bHI$}
We follow a similar procedure to obtain constraints on $\bHI$ by using an additional prior on $\OmegaHI$. To motivate a prior on $\OmegaHI$, we use measurments shown in \cref{fig:omega_HI}, except those that involve CHIME data to avoid using the same information multiple times. We use a Gaussian prior on $\OmegaHI$ centered at the fiducial value $\OmegaHI(z = 1.16)$ from \cref{eq:OmegaHIfid}, with standard deviation $\sigma_{\OmegaHI}$. We obtain the width of the prior using the scatter in the non-CHIME $\OmegaHI$ measurements. The weighted mean of these measurements is $0.54$ and the weighted standard deviation is $0.06$, where we use weights given by the inverse square of each measurement error. This provides a measure of the spread of $\OmegaHI$ values compared to the central value. We find the ratio is $0.06/0.54 \approx 0.11$. 
To be conservative, we choose a wider prior on $\OmegaHI$, with width $\sigma_{\OmegaHI}$ determined from 
$\sigma_{\OmegaHI}/\OmegaHI(z = 1.16) = 0.25$.

\begin{figure}
   \centering \includegraphics[width=\linewidth, keepaspectratio, trim = 0 0 0 0]{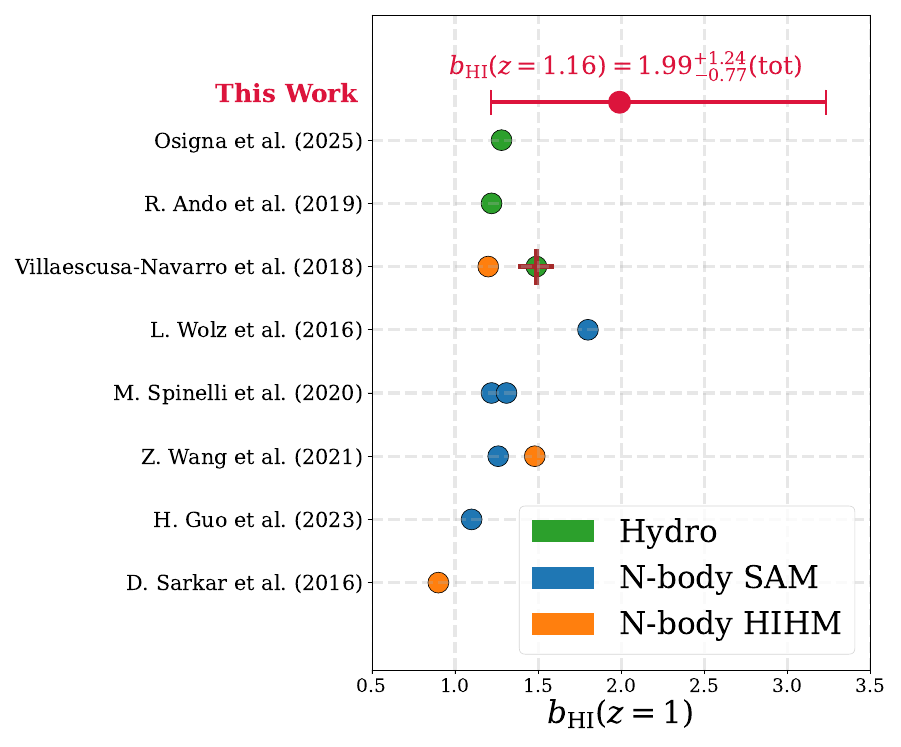}
    \caption{Constraint on $\bHI(z = 1.16)$ inferred using CHIME auto-power spectrum with an additional prior on $\OmegaHI$. We compare this estimate with the compilation of $\bHI (z = 1)$ values from various simulations, compiled in \cite{osinga2025-HImodelling}. Note the different redshifts for our constraint ($z = 1.16$) and the simulation-derived values ($z = 1$); however, based on $\bHI$ measurements from TNG100, we expect the bias to evolve by no more than 5\% between these redshifts. Simulation values are obtained using Hydro and N-body simulations, with two different ways to obtain the HI distribution: semi-analytic model (SAM) and the HI-halo mass relation (HIHM). The red `+' indicates $\bHI$ associated with our fiducial $\bHI(z)$ model based on \citealt{villaescusa-navarro2018}.
    Our $\bHI$ posterior is skewed slightly higher than, but is statistically consistent with, typical values from simulations.
    }
    \label{fig:compilation_bias_HI}
\end{figure}

In \cref{fig:compilation_bias_HI} we show our $\bHI$ constraint, $\bHIMargConstraint$, with statistical uncertainty represented as the 68\% HPDI around the mode of the distribution, combined in quadrature with the systematic uncertainty.
We compare our constraint with the bias derived from different simulation techniques, compiled in \cite{osinga2025-HImodelling}. Note that our measurement is reported at $z = 1.16$ and the values from simulations reported in the literature are at $z = 1$. The inferred $\bHI(z = 1.16)$ lies at the higher end of the $\bHI(z = 1)$ values from the simulation. This shift cannot entirely be accounted for by redshift evolution of the bias from $z=1$ to $z=1.16$: using our fiducial form of $\bHI(z)$ from \cref{eq:bHI-fitfunc} (itself based on interpolating measurements from TNG100 in \citealt{villaescusa-navarro2018}) only accounts for a 5\% difference. 
This indicates that our constraint on $\bHI$ with an additional prior on $\OmegaHI$ is skewed higher than the typical values obtained in simulations.

\section{Comparison with hydrodynamical simulations}
\label{sec:tng}

\begin{figure}[t]
   \centering \includegraphics[width=\linewidth, keepaspectratio, trim = 0 0 0 0]{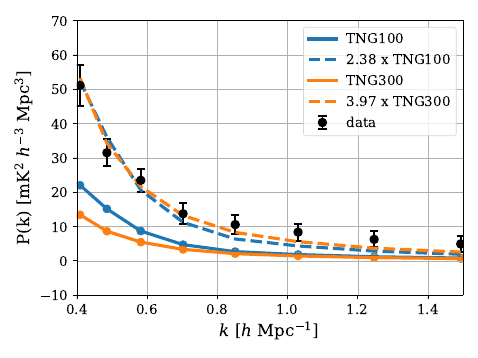}
    \caption{%
    Comparison of the CHIME auto power spectrum measurement with CHIME simulations based on IllustrisTNG-derived \tcm power spectra. Black points denote CHIME's measured bandpowers with their statistical $1\sigma$ uncertainty. Continuous curves represent the power spectrum corresponding to two Illustris-TNG boxes, TNG100 and TNG300. We fit these two power spectra to the data to obtain an amplitude parameter. The dashed curves are the continuous curves scaled by the respective amplitude parameters. After rescaling, the IllustrisTNG curves provide a good fit to the measurements, so the discrepancy can be attributed to an overall amplitude mismatch at $z\sim 1$ on nonlinear scales between the simulations and measurements.
    Accounting for statistical and systematic uncertainties, the discrepancy is at the level of $\TNGOneHundredNSigmas$ for TNG100 and $\TNGThreeHundredNSigmas$ for TNG300.}
    \label{fig:data_tng_Pk_comparison}
\end{figure}

Comparisons between the CHIME \tcm auto power spectrum and the IllustrisTNG predictions $P^{\rm TNG}_{21, \rm obs}(k)$ provide an important accuracy test for the distribution of HI in these simulations at scales probed by the data. 
\cref{fig:data_tng_Pk_comparison} shows the \tcm power spectra measured by CHIME and predicted by the two IllustrisTNG runs we consider in this work.
The TNG100 and TNG300 curves differ in both amplitude and shape. This is expected based on previous investigations demonstrating that the properties of {\HI} have not converged between the two simulations; for example, \citet{diemer2018-tng1} found that the {\HI}-mass-to-stellar-mass ratio at fixed galaxy stellar mass is a factor of $2-3$ higher in TNG100 than in TNG300, which is likely to contribute to the higher amplitude we observe for the TNG100 curve in  \cref{fig:data_tng_Pk_comparison}.

We find a significant mismatch between both IllustrisTNG predictions and the CHIME measurement. To quantify this mismatch, we allow for $P^{\rm TNG}_{21, \rm obs}(k)$ to be rescaled by a scaling parameter $A_{\rm TNG}$, and fit for separate values of this parameter for TNG100 and TNG300. 
We do so by minimizing the $\chi^2$ from \cref{eq:chi2theta} with $\vec{\theta} = [A_{\rm TNG}]$, and determining the statistical uncertainty on $A_{\rm TNG}$ by finding the boundaries of the region with $\chi^2 = \chi^2_{\rm min} + 1$. We determine the systematic uncertainty by combining the ``template sample variance", ``beam model", and ``flux scale" uncertainties from \cref{tab:AHIconstraints} in quadrature; the first two contributions affect the simulations, while the third affects the data, so all three are relevant here.

We find $A_{\rm TNG} = \TNGOneHundredAmp$ for TNG100 and $A_{\rm TNG} = \TNGThreeHundredAmp$ for TNG300.
Adding statistical and systematic uncertainties in quadrature, these correspond to $\TNGOneHundredNSigmas$ and $\TNGThreeHundredNSigmas$ discrepancies between the data and predictions from TNG100 and TNG300, respectively. 
These numbers are systematics-limited; assuming zero systematic uncertainty would instead yield $7.7\sigma$ and $9.9\sigma$, respectively.

The disagreement between simulation and the data is at the level of overall amplitude, in the sense that each scaled version of $P^{\rm TNG}_{21, \rm obs}(k)$ provides a good fit to the data, with comparisons between the data and $A_{\rm TNG} P^{\rm TNG}_{21, \rm obs}(k)$ at the best fit $A_{\rm TNG}$ values yielding $\chi^2=9.88$ for TNG100 and $4.51$ for TNG300. The corresponding $p$-values (for $7$ degrees of freedom) are $0.19$ and $0.72$, indicating reasonable fits.

This discrepancy points to a significant difference between the clustering of {\HI} at the scales measured by CHIME and that produced in the IllustrisTNG simulations. The mean {\HI} density $\OmegaHI$ is unlikely to explain this difference on its own, since the model-based CHIME constraint on $\OmegaHI$ in \cref{fig:omega_HI} is consistent with the simulation values ($6.65\times 10^{-4}$ for TNG100 and $6.25\times 10^{-4}$ for TNG300, after rescaling the $z=1$ measurements to $z=1.16$; see Appendix~\ref{app:tng:rescaling}), and the simulation values were also found to be consistent with other $\OmegaHI$ estimates in \citet{villaescusa-navarro2018}. 

There are several other aspects of the simulations that affect the {\HI} power spectrum, including the amount of {\HI} in halos of different masses, the occupation of {\HI} in central versus satellite galaxies as a function of halo mass, the distribution of {\HI} in these galaxies, and the velocity dispersion of {\HI} on different scales. 
None of these properties were calibrated in advance; instead, they are indirectly determined by the numerical and modelling choices in the simulations, including the implementations of star formation, feedback from supernovae and active galactic nuclei, and ultraviolet background radiation. \tcm power spectrum measurements therefore represent a new constraint on these prescriptions, and further effort will be required to determine the precise reason(s) for the discrepancy with CHIME's observations.

Other discrepancies between IllustrisTNG's {\HI} predictions and observations have also been noted. \citet{diemer2019-tng2} found that the {\HI} density in TNG100 at $z=0$ is roughly twice that inferred from direct \tcm galaxy surveys, and contains a larger-than-expected population of satellite galaxies that are extremely gas-poor.
\citet{stevens2019-tng} found that stripping of {\HI} from satellite galaxies (via ram pressure or tidal effects) happens concurrently with those galaxies becoming quenched, contrary to expectations of a time delay between stripping and quenching. It remains to be seen whether these properties of the simulations are related to the power spectrum discrepancy we have observed.

\section{Conclusions}
\label{sec:conclusions}

\citet{chime-auto-paper} has presented a measurement of the \tcm auto power spectrum at $z\sim 1$ using the Canadian Hydrogen Intensity Mapping Experiment, in the frequency band from \SIrange{608.2}{707.8}{\mega\hertz} and in the wavenumber range $0.4\ihMpc \lesssim k \lesssim 1.5\ihMpc$. In this paper, we have presented the results of two different approaches to interpreting this measurement.

In the first approach, we have fit a 4-parameter model to the power spectrum, and determined the posteriors on these parameters using a Bayesian analysis. This model involves the {\HI} density parameter $\OmegaHI$, the {\HI} bias $\bHI$, and two phenomenological parameters describing the shape of the power spectrum: $\alphaNL$, which interpolates between linear and nonlinear models for the matter power spectrum, and $\alphaFoG$, which controls the severity of Finger-of-God damping of line-of-sight modes based on an approximate model for this damping. In our main results, we marginalize over the two shape parameters and constrain a degenerate combination of~$\OmegaHI$ and~$\bHI$, defined by $\AHIps = 10^6 \OmegaHI^2(\bHI^2+\fmu)^2$, at the mean redshift of the measurement ($z=1.16$), obtaining
$\AHIpsConstraintFullBand$. We also present separate constraints on $\OmegaHI$ and $\bHI$ if additional prior information is assumed about one of the two; this yields 
$\OmegaHIMargConstraint$
and
$\bHIMargConstraint$.
This constraint on $\OmegaHI$ is slightly higher than, but mostly consistent with, estimates based on \tcm stacking, damped Lyman-$\alpha$ systems, and previous intensity mapping cross-correlations. Most simulation-based estimates of $\bHI$ at $z=1$ fall in the range $1\lesssim \bHI \lesssim 1.5$, and our constraint on $\bHI(z=1.16)$ is mostly consistent with these values but also extends to much higher values.

In the second approach, we compare the measured power spectrum with {\HI} power spectra from the TNG100 and TNG300 runs of the IllustrisTNG suite of simulations, by propagating power spectra measured from the simulations through the CHIME simulation and analysis pipelines.
We find that TNG100 and TNG300 are discrepant with the CHIME measurements at $\TNGOneHundredNSigmas$ and $\TNGThreeHundredNSigmas$, respectively, with the discrepancy mostly attributable to a mismatch in overall power spectrum amplitude (as opposed to scale-dependence) at the relevant scales.
The value of the mean {\HI} density $\OmegaHI$ is unlikely to be the sole explanation for this discrepancy; instead, we conclude that the simulations predict much weaker redshift-space clustering of {\HI} at $0.4\ihMpc \lesssim k \lesssim 1.5\ihMpc$ than indicated by our observations.
It remains an open question to determine which ingredients of the simulations are responsible for the discrepancy, and what this can teach us about the astrophysical processes involved.

There are several additional possibilities for extracting astrophysical information from the measurement in \citet{chime-auto-paper}
that we have not explored in this work. Power spectrum predictions can be derived from {\HI} halo models and used to learn about how {\HI} is distributed amongst dark matter halos and galaxies with different properties \citep[e.g.][]{padmanabhan2017-halomodel,wolz2019-halomodel,chen2021-halomodel,schaan2021-LIMhalomodel}. Semi-analytical models built atop $N$-body simulations allow for nonlinear structure formation to be captured in more detail, while retaining flexibility in how different astrophysical phenomena and properties are modelled \citep[e.g.][]{wolz2016-sam,spinelli2020-sam,li2024-sam}. Further comparisons with other hydrodynamical simulations will also be informative to assess the realism of the subgrid algorithms they contain, especially in light of recent indications from Sunyaev-Zeldovich observations that feedback processes in some simulations are too weak to explain the inferred gas profiles \citep[e.g.][]{bigwood2024-szfeedback,hadzhiyska2025-szfeedback,pandey2025-szfeedback,sunseri2025-szfeedback,siegel2025-szfeedback,mccarthy2025-szfeedback}.

Future CHIME measurements will also yield improved constraints on nonlinear {\HI} clustering. Improved data cleaning and analysis techniques will enable inclusion of more observation time and a wider portion of the frequency band, opening up the possibility of measuring redshift evolution of the power spectrum. Larger-scale (lower-$k$) measurements will enhance our ability to disentangle different scale-dependent effects, such as redshift-space distortions. Joint analyses between auto spectra and cross-correlations with galaxies \citep{chimestacking}, the Lyman-$\alpha$ forest \citep{chime-lymanalpha}, gravitational lensing \citep{chime-lensing-bispectrum}, or other tracers of large-scale structure will elucidate the behavior of {\HI} in the environments of these tracers. Beyond-two-point statistics, such as the bispectrum, can also provide distinct information from the power spectrum due to their ability to break parameter degeneracies in models of baryonic physics \citep{foreman-matter-bispectrum,yankelevich-halo-bispectrum}.

Our interpretation methods in this work were chosen partially because they do not require explicit knowledge of the transfer function that relates a prediction for the cosmological \tcm power spectrum to the measured spectrum. Work to characterize this transfer function is ongoing within the CHIME collaboration, and will allow for much more flexibility in the form of parametric models that can be constrained.

Work is also underway to improve CHIME's control of systematics and foregrounds, to enable access to lower-$k$ modes where cosmological information can be obtained from baryon acoustic oscillations and perturbative modelling. We expect \tcm intensity mapping measurements from CHIME and other telescopes to play an increasing role in the ongoing endeavor to learn more about the fundamental laws of our universe. However, in this paper we have demonstrated that astrophysical information can be extracted from intensity mapping at nonlinear scales, and we look forward to future insights from these scales in the years to come.


\section*{acknowledgments}

We thank the Dominion Radio Astrophysical Observatory, operated by the National Research Council Canada, for gracious hospitality and expertise. The DRAO is situated on the traditional, ancestral, and unceded territory of the syilx Okanagan people. We are fortunate to live and work on these lands.

CHIME is funded by grants from the Canada Foundation for Innovation (CFI) 2012 Leading Edge Fund (Project 31170), the CFI 2015 Innovation Fund (Project 33213), and by contributions from the provinces of British Columbia, Qu\'ebec, and Ontario. Long-term data storage and computational support for analysis is provided by WestGrid\footnote{\url{https://www.westgrid.ca/}}, SciNet\footnote{\url{https://www.scinethpc.ca/}} and the Digital Research Alliance of Canada \footnote{\url{https://www.alliancecan.ca/}}, and we thank their staff for flexibility and technical expertise that has been essential to this work, particularly Martin Siegert, Lixin Liu, and Lance Couture.

Additional support was provided by the University of British Columbia, McGill University, and the University of Toronto. CHIME also benefits from NSERC Discovery Grants to several researchers, funding from the Canadian Institute for Advanced Research (CIFAR), and from the Dunlap Institute for Astronomy and Astrophysics at the University of Toronto. The Dunlap Institute is funded through an endowment established by the David Dunlap family and the University of Toronto.

This material is partly based on work supported by the NSF through grants 2006911, 2008031,   2510770, 2510771, 2510772, and 2510773, and by the Perimeter Institute for Theoretical Physics, which in turn is supported by the Government of Canada through Industry Canada and by the Province of Ontario through the Ministry of Research and Innovation. 
We acknowledge the support of the Natural Sciences and Engineering Research Council of Canada (NSERC) [funding reference number 569654].
M.~D. is supported by a CRC Chair, NSERC Discovery Grant, CIFAR, and by the FRQNT Centre de Recherche en Astrophysique du Qu\'ebec (CRAQ).
K.~W.~M. holds the Adam J.~Burgasser Chair in Astrophysics.
J.~M.~P. acknowledges the support of an NSERC Discovery Grant (RGPIN-2023-05373).
We thank the Beus Center for Cosmic Foundations at Arizona State University for supporting a workshop where part of this work was completed.

\software{
   angpow \citep{campagne2017-angpow},
   CAMB \citep{lewis1999,camb_zenodo},
   caput \citep{caput},
   ch\_pipeline \citep{ch_pipeline},
   cora \citep{cora},
   Cython \citep{Cython},
   draco \citep{draco},
   driftscan \citep{driftscan},
   emcee \citep{foreman-mackey2013},
   FFTW \citep{FFTW05},
   GetDist \citep{getdist-jcap,getdist_zenodo},
   h5py \citep{h5py},
   hankl \citep{karamanis2021},
   HDF5 \citep{HDF5},
   healpy \citep{healpy},
   Matplotlib \citep{Matplotlib,matplotlib_zenodo},
   mpi4py \citep{mpi4py_2021,mpi4py_2023},
   NumPy \citep{NumPy},
   OpenMPI \citep{OpenMPI},
   SciPy \citep{SciPy}
   Skyfield \citep{Skyfield},
   zarr \citep{zarr_zenodo}.
   }

\appendix
\twocolumngrid

\section{Details of model-evaluation pipeline}
\label{app:model_details}

In this appendix, we provide further details about the procedure for evaluating the power spectrum model described in \secref{sec:modelling}.

\subsection{Pipeline linearity}
\label{app:model_details:linearity}

In this subsection, we argue for the validity of forming a linear combination of pre-computed power spectrum templates as a means to capture the model's dependence on $\alpha_\Omega$, $\alpha_b$, and $\alphaNL$, where \cref{eq:OmegaHIfitpar,eq:bHI-fitpar} define the relationship between the first two parameters and $\OmegaHI$ and $\bHI$.

Let $m_i$ represent a pixelized sky map of \tcm fluctuations, with $i$ indexing angular pixel and frequency, such that the underlying cosmological correlation function can be written as
\beq
\xi_{ii'} = \left\langle m_i m_{i'} \right\rangle\ .
\eeq
Similarly, let $d_j$ represent a vector of ringmap values.
If the relationship between the sky map and ringmap (including intermediate steps that relate the sky map to observed visibilities and the visibilities to a filtered, masked ringmap) is linear, we can write this relationship using a transfer function $T_{ji}^{\rm (map)}$ as
\beq
d_j = \sum_i T_{ji}^{\rm (map)} m_i\ .
\label{eq:linearity_dj}
\eeq
If the pipeline for measuring a power spectrum consists of correlating the (Fourier-transformed) map with itself and applying further operations that are linear in the squared map, then it can be represented as
\beq
P_k = \sum_{j} T_{kj}^{\rm (power)} |d_j|^2\ .
\label{eq:linearity_Pk}
\eeq
When analyzing CHIME data, we take two splits of the data and perform a cross-correlation, but here we are only concerned with the \tcm contribution, so we assume that each split contains the same \tcm realization.

We can use the above setup to justify our treatment of the 3 parameters mentioned earlier. Starting with $\alpha_\Omega$, we note that
our model for cosmological \tcm fluctuations implies that a sky map $m_i$ is directly proportional to this parameter.
Therefore, \cref{eq:linearity_dj} and \cref{eq:linearity_Pk} imply that the observed power spectrum is proportional to $\alpha_\Omega^2$.

To discuss $\alpha_b$, we write our sky map to factor out the dependence on the {\HI} bias and Kaiser contribution:
\beq
m_i = \lp
	\alpha_b\, b_i + f_i \mu_i^2
\rp \tilde{m}_i \ .
\eeq
Propagating this through
\cref{eq:linearity_dj} and \cref{eq:linearity_Pk}, we find that
\beq
P_k(\alpha_b) =
	\alpha_b^2 P_k^{\rm hh} + \alpha_b P_k^{\rm hv} + P_k^{\rm vv}\ ,
\label{eq:Pk_b}
\eeq
where
\begin{align*}
\numberthis
P_k^{\rm hh} &= 
	\sum_j T_{kj}^{\rm (power)}
	\sum_{i,i'} T_{ji}^{\rm (map)} T_{ji'}^{{\rm (map)}*}
	\tilde{m}_i \tilde{m}_{i'}\ , \\
P_k^{\rm hv} &= 
	2 \sum_j T_{kj}^{\rm (power)} \\
\numberthis
&\qquad\times
	\sum_{i,i'} T_{ji}^{\rm (map)} T_{ji'}^{{\rm (map)}*}
	f_i \mu_i^2 \tilde{m}_i \tilde{m}_{i'}\ , \\
P_k^{\rm vv} &= 
	\sum_j T_{kj}^{\rm (power)} \\
\numberthis
&\qquad\times
	\sum_{i,i'} T_{ji}^{\rm (map)} T_{ji'}^{{\rm (map)}*} 
	f_i f_{i'} \mu_i^2 \mu_{i'}^2 \tilde{m}_i \tilde{m}_{i'}\ .
\end{align*}
Therefore, if we can produce simulated maps with either (1) $\alpha_b=0$ with the Kaiser term or (2) $\alpha_b=1$ and no Kaiser term, the analysis pipeline can be used to generate the three terms in the equations above, and then we can assemble them into \cref{eq:Pk_b} with the appropriate prefactors of~$\alpha_b$. In practice, we do not generate these exact maps, but instead generate other maps and then take linear combinations of the resulting power spectra that reproduce the 3 terms we require. This process is described in Appendix~\ref{app:model_details:template_combinations}.

For $\alphaNL$, on the other hand, we note that the sky maps do not depend linearly on this parameter, but instead it is the underlying cosmological power spectrum that has the linear dependence (see \cref{eq:Pm-model}). Therefore, the corresponding correlation function has the same dependence:
\beq
\xi_{ii'}(\alphaNL) = \alphaNL \xi_{ii'}^{\rm NL} + \lp 1-\alphaNL \rp \xi_{ii'}^{\rm L}\ .
\eeq
By using this expression when taking an expectation value of
\cref{eq:linearity_Pk}, we find
\beq
\left\langle P_k(\alphaNL) \right\rangle
	= \alphaNL \left\langle P_k^{\rm NL} \right\rangle
	+ \lp 1-\alphaNL \rp \left\langle P_k^{\rm L} \right\rangle\ ,
\label{eq:Pk_alphaNL}
\eeq
where
\beq
\left\langle P_k^{\rm NL} \right\rangle
	= \sum_j T_{kj}^{\rm (power)}
	\sum_{i,i'} T_{ji}^{\rm (map)} T_{ji'}^{{\rm (map)}*}
	\xi_{ii'}^{\rm NL}
\eeq
and similarly for the ``${\rm L}$" term. \cref{eq:Pk_alphaNL} shows that we can generate separate simulations that use linear and nonlinear input matter power spectra, and take a linear combination of the resulting output spectra to evaluate the power spectrum model for arbitrary values of $\alphaNL$.

Since each argument above is related to linearity in one of the 3 parameters, the arguments can be combined to justify the linear form of the model, shown in \cref{eq:Pobs-6terms} in the next subsection. The Finger-of-God parameter $\alphaFoG$ cannot be handled in this way, and we describe a separate scheme for this parameter in Appendix~\ref{app:model_details:fog}.

We note that there are several steps of the data processing pipeline in \citet{chime-auto-paper} that are nonlinear in visibilities: RFI flagging, bandpass gain corrections using the HyFoReS algorithm \citep{hyfores_2022,hyfores_2025a,hyfores_2025b}, noise cross-talk removal, and masking of \tcm absorption systems. 
However, our RFI flagging algorithms and \tcm absorber masking procedures identify excursions in the data that are much larger than the expected \tcm fluctuations, so they will not introduce significant nonlinearity in the \tcm fluctuations themselves. 
Noise cross-talk removal is only nonlinear because of the usage of a median over a narrow range of Earth rotation angle ($\Delta{\rm ERA} = 15^\circ$). Regarding HyFoReS, we have verified that the power spectrum measurement with and without using this algorithm is consistent within the uncertainties on the measurement.
Therefore, we do not expect the small amount of nonlinearity introduced by these steps to significantly affect the arguments in this subsection.

\subsection{Template combinations}
\label{app:model_details:template_combinations}

\cref{eq:P21theory_sum} displays our \tcm power spectrum model as a sum of 6 terms. As argued in the previous subsection, after applying the transfer function for our measurement pipeline, the model can be written as
\begin{align*}
&P_{\rm obs}(\alpha_\Omega, \alpha_b, \alphaNL) \\
&\quad= \alpha_\Omega^2 \alpha_b^2 (1-\alphaNL) P_{\rm obs}^{\rm hh,L}
+ \alpha_\Omega^2 \alpha_b^2 \alphaNL P_{\rm obs}^{\rm hh,NL} \\
&\quad\quad
	+ \alpha_\Omega^2 \alpha_b (1-\alphaNL) P_{\rm obs}^{\rm hv,L} 
	+ \alpha_\Omega^2 \alpha_b \alphaNL P_{\rm obs}^{\rm hv,NL} \\
&\quad\quad
	+ \alpha_\Omega^2  (1-\alphaNL) P_{\rm obs}^{\rm vv,L} 
	+ \alpha_\Omega^2 \alphaNL P_{\rm obs}^{\rm vv,NL} \ ,
	\numberthis
\label{eq:Pobs-6terms}
\end{align*}
omitting the dependence on $k$, $\mu$, $z$, and $\alphaFoG$ for brevity. Therefore, we would like to generate simulations that allow us to pre-compute each of the $P$ terms on the right-hand side. To do so, we make use of the fact that each of these terms can be written as a linear combination of evaluations of the full model $P_{\rm obs}(\alpha_\Omega, \alpha_b, \alphaNL)$ at specific values of $\alpha_\Omega$, $\alpha_b$, and $\alphaNL$; specifically,
\begin{align}
\nonumber
P_{\rm obs}^{\rm hh,L}
	&= 2P_{\rm obs}(1, 1, 0) - 4P_{\rm obs}\!\lp  1, \frac{1}{2}, 0 \rp  \\
\label{eq:Pobs_lc1}
&\quad
	+ 2P_{\rm obs}(1, 0, 0)\ , \\
\nonumber
P_{\rm obs}^{\rm hv,L}
	&= P_{\rm obs}\!\lp 1, \frac{1}{2}, 0 \rp - P_{\rm obs}(1, 0, 0)  \\
\label{eq:Pobs_lc2}
&\quad
	- \frac{1}{4} P_{\rm obs}^{\rm hh,L}\ , \\
\label{eq:Pobs_lc3}
P_{\rm obs}^{\rm vv,L}
	&= P_{\rm obs}(1, 0, 0)\ ,
\end{align}
and similarly for the ``NL" templates if each term is evaluated with $\alphaNL=1$ instead of $0$.

To generate the 6 terms in \eqref{eq:Pobs-6terms}, we generate 6 simulations with input model parameters
\begin{align*}
&(\alpha_\Omega, \alpha_b, \alphaNL) \\
&\qquad \in
    \left\{
    	(1, 0, 0), \lp 1, \frac{1}{2}, 0 \rp, (1, 1, 0), 
    \right. \\
&\qquad\quad\;\;\;
    \left.
    	(1, 0, 1), \lp 1, \frac{1}{2}, 1 \rp, (1, 1, 1)
    \right\}\ ,
    \numberthis
    \label{eq:3alpha-values}
\end{align*}
measure their power spectra, and form the linear combinations in \cref{eq:Pobs_lc1} to \cref{eq:Pobs_lc3}.

\subsection{Finger-of-God parameter interpolation}
\label{app:model_details:fog}

The dependence of our power spectrum model on the $\alphaFoG$ parameter is more complex than the other 3 parameters (which each enter the model linearly or quadratically), so this parameter must be handled separately. We do so by interpolating between templates that have been pre-computed for a range of $\alphaFoG$ values from $0$ to $20$: specifically, 
\begin{align*}
\alphaFoG &\in \left\{
	0,0.1,0.15,0.2,0.3,0.4,0.6,0.8,1, \right. \\
&\quad\;
\left.
	1.2,1.5,2,3,4,5,6,8,10,15,20
\right\}\ .
\numberthis
\end{align*}

For each of these values, we generate simulations with the 6 parameter combinations in \cref{eq:3alpha-values}, and form the linear combinations in \cref{eq:Pobs_lc1} to \cref{eq:Pobs_lc3}. For each linear combination, indexed by $Z$, the lowest-$k$ power spectrum value is well-described by
\begin{align*}
&P_{\rm obs}^Z(k_{\rm min};\alphaFoG) \\
&\qquad = r_Z(\alphaFoG) P_{\rm obs}^Z(k_{\rm min};\alphaFoG=1)
\numberthis
\label{eq:Pobs_rscaling}
\end{align*}
with
\beq
r_Z(\alphaFoG) \equiv \frac{(1+c_Z)^2}{(1+c_Z \alphaFoG^2)^2}\ ,
\label{eq:rZ}
\eeq
where each $c_Z$ is an order-unity constant that we fit to the pre-computed templates. This function is motivated by the Finger-of-God damping kernel in our cosmological power spectrum model, given by \cref{eq:DFoGk}; while the pipeline transfer function modifies the dependence of the observed power spectrum on $\alphaFoG$, \cref{eq:rZ} captures the majority of this dependence. To describe the residual dependence on $\alphaFoG$, we compute the ratio of the left- and right-hand sides of \cref{eq:Pobs_rscaling} at each $k$, and fit 
cubic splines $s_Z$ in $\alphaFoG$ to those ratios, resulting in one spline per $Z$ and $k$:
\beq
s_Z(k;\alphaFoG) 
	= \frac{
		P_{\rm obs}^Z(k;\alphaFoG)
	}{
		r_Z(\alphaFoG) P_{\rm obs}^Z(k;\alphaFoG=1)		
	}\ .
\eeq

When evaluating the full power spectrum model, the pre-computed $c_Z$ constants and $s_Z$ splines are used to evaluate each $P_{\rm obs}^Z$ term at the desired value of $\alphaFoG$,
\begin{align*}
P_{\rm obs}^Z(k; \alphaFoG) &= 
	s_Z(k; \alphaFoG) r_Z(\alphaFoG) \\
&\quad
	\times P_{\rm obs}^Z(k; \alphaFoG=1)\ .
	\numberthis
\end{align*}
These terms are then assembled into the form in \cref{eq:Pobs-6terms}.


\section{Details of simulation pipeline}
\label{app:sim_details}

In this appendix, we provide further details of the sky simulations described in \secref{sec:simulations:method}.

\subsection{Correlation functions}

We transform an input matter power spectrum, evaluated at $z=1$, into real-space correlation functions via
\beq
\xi_{AB}(r) = \frac{1}{2\pi^2} \int dk\, k^2 \frac{\sin kr}{kr} k^{n_A+n_B} P_{\rm m}(k)\ ,
\label{eq:xiAB}
\eeq
where $A$ and $B$ refer to the matter overdensity $\deltam$ or a rescaled version of the gravitational potential\footnote{If we write the Poisson equation for the gravitational potential $\Phi$ as $k^2\Phi(\vk,a) = (3/2)\Omega_{\rm m} H_0^2 a^{-1} \deltam(\vk,a)$, then we define $\phi(\vk,a) \equiv (3/2)\Omega_{\rm m} H_0^2 a^{-1}\Phi(\vk,a)$, so that $k^2\phi(\vk,a) = \deltam(\vk,a)$.}, which we call $\phi$.
In \cref{eq:xiAB}, the powers of $k$ in the integral are defined by $n_{\deltam}=0\, , n_\phi=2$.
For $r< 10\hMpc$, we evaluate \cref{eq:xiAB} using Romberg integration in $\log k$, while at higher $r$, we use the FFTlog method \citep{hamilton2000-fftlog} as implemented in the \texttt{hankl} package \citep{karamanis2021}, using Richardson extrapolation in $\Delta\log k$ to improve convergence. We compute each correlation function on a log-spaced grid in $r$, and construct a cubic spline in a $\sinh$-transformed space that effectively interpolates in log while smoothly handling negative values.

\subsection{Multi-frequency angular power spectra}
\label{app:sim_details:aps}

\subsubsection{Overview}

Correlation functions defined in \cref{eq:xiAB} can be transformed into multi-distance angular power spectra by evaluating
\begin{align*}
&C_\ell^{AB}(\chi_1, \chi_2)  \\
&\qquad= 2\pi \int_{-1}^1 d\mu\, \mathcal{P}_\ell(\mu)\, 
	\xi_{AB}\!\lp r[\chi_1,\chi_2,\mu]\rp\ ,
\numberthis
\label{eq:CellAB}
\end{align*}
where $r[\chi_1,\chi_2,\mu] \equiv \sqrt{\chi_1^2+\chi_2^2-2\chi_1\chi_2\mu}$ and $\mathcal{P}_\ell(\mu)$ is the Legendre polynomial of degree $\ell$. 
However, the results will be in terms of $\deltam$ and $\phi$ instead of $\deltam$ and the Kaiser term, and will not incorporate Finger-of-God damping or the appropriate integration over CHIME's frequency channel profile. These features can be captured by modifying \cref{eq:CellAB} like so:
\begin{align*}
C_{\ell f_1 f_2}^{AB}
	&= 2\pi \int_{-1}^1 d\mu\, \mathcal{P}_\ell(\mu) \\
&\quad\times \int d\chi_1\, W_{f_1}(\chi_1) 
	\int d\chi_2\, W_{f_2}(\chi_2) \\
&\quad\times 
	\int d\chi_1'\, D_\sHI^{\rm FoG}(\chi_1 - \chi_1')
	\int d\chi_2'\, D_\sHI^{\rm FoG}(\chi_2 - \chi_2') \\
&\quad\times
	\lp \frac{\partial}{\partial\chi_1'} \rp^{n_A}
	\lp \frac{\partial}{\partial\chi_2'} \rp^{n_B}
	\xi_{AB}\!\lp r[\chi_1',\chi_2',\mu]\rp
\numberthis
\label{eq:Clf1f2AB}
\end{align*}
We evaluate the $\mu$ integral using Gauss-Legendre quadrature. 
In the following subsections, we describe how we evaluate the integrand for each $\mu$ value.

\subsubsection{Kaiser contribution}

We sample $\xi_{AB}\!\lp r[\chi_1',\chi_2',\mu]\rp$ on a uniform grid in $(\chi_1',\chi_2')$, bounded by the minimum and maximum comoving line-of-sight distances contained with the relevant frequency band, plus at least one extra frequency channel on either side (with the precise number of extra channels determined by the Finger-of-God scheme described later). If $A$ or $B=\phi$, we compute second derivatives in $\chi_1'$ or $\chi_2'$ via finite differences, which converts that leg of the correlation function from $\phi$ to the Kaiser-term field $\mu^2\deltam$. By ensuring that the set of frequency channels is padded by at least one extra channel, we are able to
use the same two-sided finite-difference computation across the entire relevant $\chi$ range, without the need to use one-sided differences at each edge.

\subsubsection{Finger-of-God damping}

Our power spectrum model implements Finger-of-God damping using the Lorentzian factor in \cref{eq:DFoGk}. This Fourier-space damping is equivalent to a (fixed-redshift) convolution in line-of-sight distance with the following kernel \citep[e.g.][]{scoccimarro2004}:
\begin{align*}
&D_\sHI^{\rm FoG}(\Delta\chi, z; \alphaFoG) \\
&\qquad= \frac{1}{\sqrt{2}\, \alphaFoG\, \sigma_{\rm FoG}^{\rm (fid)}(z)} \\
&\qquad\quad\times
	\exp\!\lp -\frac{\sqrt{2}}{\alphaFoG\, \sigma_{\rm FoG}^{\rm (fid)}(z)} |\Delta\chi| \rp\ .
\numberthis
\label{eq:DFoGx}
\end{align*}
We simulate sky maps that incorporate this damping by convolving the input multi-distance angular power spectrum with this kernel in each $\chi$ argument, as written in the third line of \cref{eq:Clf1f2AB} (where we omit the $z$ and $\alpha_{\rm FoG}$ arguments for brevity).
To limit the computational cost of the convolution, we symmetrically truncate the $\chi$ range of the kernel such that it covers 75\% of the integral of the un-truncated function. (Our tests in Appendix~\ref{app:sim_details:validation:flat_sky} verify that this choice is sufficient for our desired accuracy.) 
We then normalize the truncated kernel to integrate to unity, such that the convolution obeys conservation of mass.
We perform the convolution over a $\chi$ range that is padded to ensure that the kernel is not cut off for points at the edges of the desired output range.

\subsubsection{Frequency channel profile}

\begin{figure}[t]
   \centering \includegraphics[width=\linewidth, keepaspectratio, trim = 0 0 0 0]{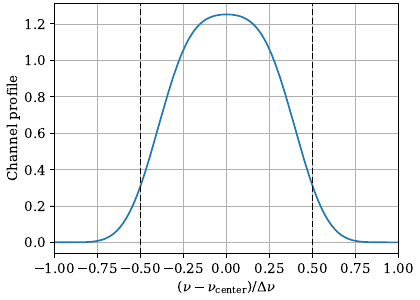}
    \caption{%
    The profile associated with visibilities measured within each CHIME frequency channel, as a function of the frequency difference from the channel's central frequency, relative to the nominal channel width of $\Delta\nu = 390\,{\rm kHz}$. The dashed vertical lines illustrate that the profile extends slightly beyond this width. We incorporate this profile into the multi-frequency angular power spectra used to generate sky simulations, such that each sky map includes the effect of integrating over the profile.
    }
    \label{fig:channel_profile}
\end{figure}

CHIME uses a polyphase filter bank (PFB) to transform digitized timestreams into $390\,{\rm kHz}$-wide frequency channels \citep{CHIMEoverview}. This procedure implies that each visibility measures the integral of the sky signal over the frequency range of each channel, weighted by a profile given by the squared Fourier transform of the PFB's time-domain window function. This profile is shown in \cref{fig:channel_profile}.

We incorporate this effect in our simulations by integrating the multi-distance angular power spectrum against the channel profile in each distance argument, as written in the second line of \cref{eq:Clf1f2AB}. We perform these integrals using Simpson's rule. \cref{fig:channel_profile} shows that the channel profile extends beyond the nominal edges of each channel; in order to capture the contribution from these tails of the profile, we integrate over the set of 3 consecutive channels containing the channel of interest.

The profile shown in \cref{fig:channel_profile} is defined as a function of $(\nu-\nu_{\rm center})/\Delta\nu$, the frequency difference from the central frequency relative to the nominal channel width. If we refer to this function as $\tilde{W}$, we define the profiles in \cref{eq:Clf1f2AB} by
\beq
W_f(\chi) = \frac{1}{\Delta\chi_f} \tilde{W}\!\lp \frac{\chi-\chi_f}{\Delta\chi_f} - \frac{1}{2} \rp\ ,
\eeq
where $\chi_f$ and $\Delta\chi_f$ are the central comoving distance of the channel and the width of the channel in comoving distance, respectively. This implements an approximation for the exact change of variables from $\nu$ to $\chi$, which is accurate to $\mathcal{O}(0.1\%)$ for the channels we consider in this work.

\subsection{Sky maps}

We use the multi-frequency angular power spectra discussed above to generate Gaussian HEALPix maps of $\deltam$ and $\mu^2\deltam$ for each frequency channel, and scale each map to the central redshift of each channel using the linear growth factor. We then scale the $\deltam$ map by the {\HI} bias, evaluated using our fiducial model from \cref{eq:bHI-fitfunc} at the central redshift of each channel. We similarly scale the $\mu^2\deltam$ map by $f(z)$, and add the result to the scaled $\deltam$ map.

\subsection{Validation}
\label{app:sim_details:validation}

\subsubsection{Comparison with external angular power spectrum code}
\label{app:sim_details:validation:angpow}

\begin{figure}[t]
   \centering \includegraphics[width=\linewidth, keepaspectratio, trim = 0 0 0 0]{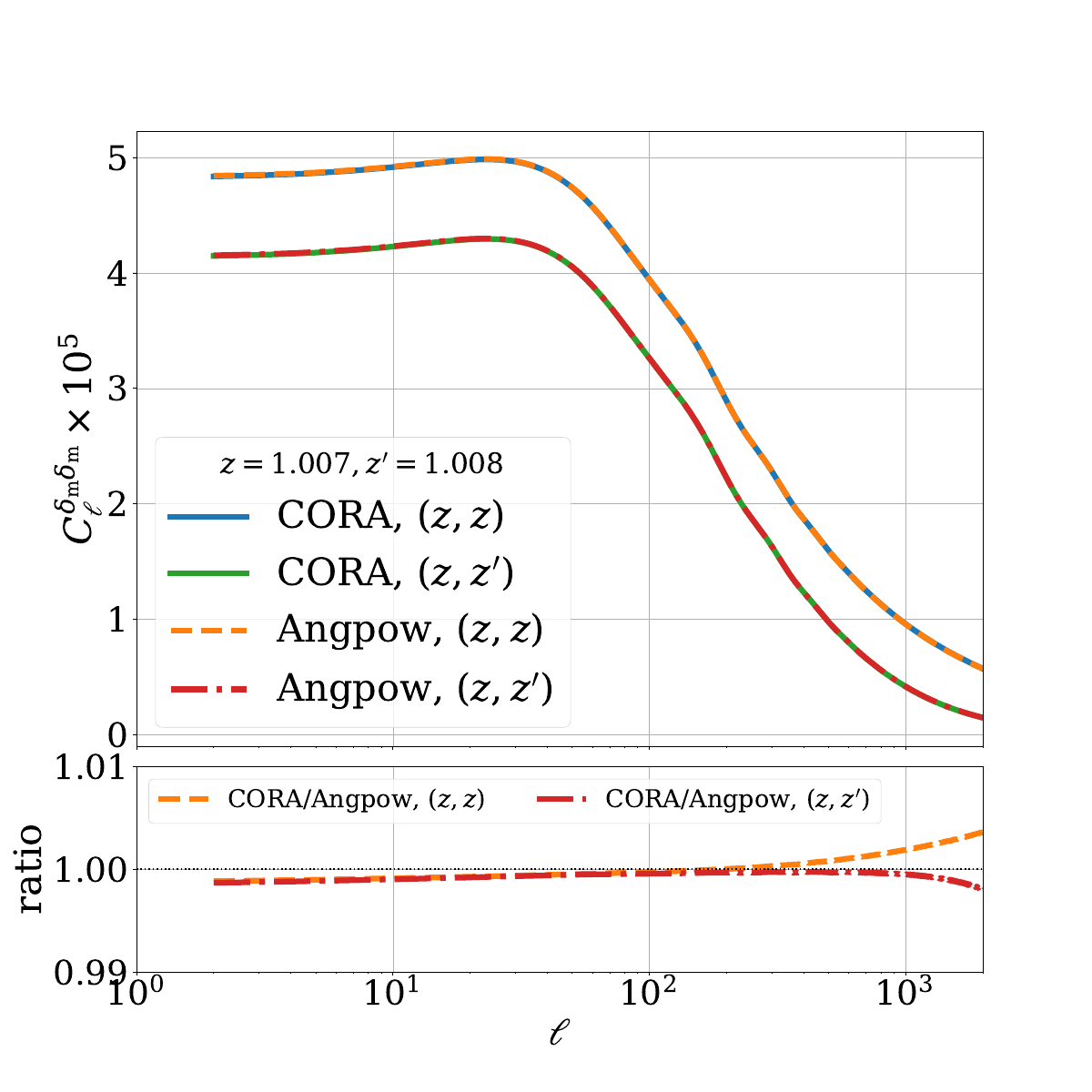}
    \caption{%
    Comparison of the angular power spectrum of $\deltam$ computed using our implementation (contained in the \texttt{cora} package) with the output from a public package, \texttt{angpow}, which computes theoretical spectra for $\deltam$ for a given window function and cosmology. We see a subpercent level agreement between the two, as indicated in the bottom panel. Two example spectra are shown, one of auto-spectrum $(z,z)$ at $z = 1.007$ and other for the cross-spectrum between nearby shells $(z,z')$ with $z' = 1.008$.}
    \label{fig:cora_angpow_comparison}
\end{figure}

In \cref{fig:cora_angpow_comparison}, we show the output of \cref{eq:Clf1f2AB} for the $\deltam$ auto power spectrum without Finger-of-God damping and using a top-hat frequency channel profile, and compare with the output of the \texttt{angpow} package \citep{campagne2017-angpow} for the same quantity. We find sub-percent agreement at the scales relevant for this work.

\subsubsection{Comparison with flat-sky approach}
\label{app:sim_details:validation:flat_sky}

We additionally validate the output of \cref{eq:Clf1f2AB} by comparing to an alternative method based on the flat-sky approximation. In this method, we first evaluate a 3d power spectrum $P(k_\parallel,k_\perp)$ directly, and then convert to a multi-frequency angular power spectrum using the flat-sky expression from \citet{datta2007-flatskyCl}, incorporating the frequency channel profile similarly to \cref{eq:Clf1f2AB}:
\begin{align*}
C_{\ell f_1 f_2} &=
	\frac{1}{\pi} \int d\chi_1\, W_{f_1}(\chi_1) 
	\int d\chi_2\, W_{f_2}(\chi_2) \\
&\quad\times
	\frac{1}{\bar{\chi}^2}
	\int dk_\parallel  \cos(k_\parallel \Delta\chi) \\
&\quad\times
	P\!\lp k_\parallel, k_\perp = \frac{\ell}{\bar{\chi}}; z_{\rm eff} \rp\ ,
\numberthis
\label{eq:Cell_flatsky}
\end{align*}
where $\Delta\chi \equiv \chi-\chi'$ and $\bar{\chi} \equiv (\chi+\chi')/2$. We evaluate the $k_\parallel$ integral with a discrete cosine transform and the $\chi$ integrals with Gauss-Legendre quadrature.

\begin{figure}[t]
   \centering \includegraphics[width=\linewidth, keepaspectratio, trim = 0 0 0 0]{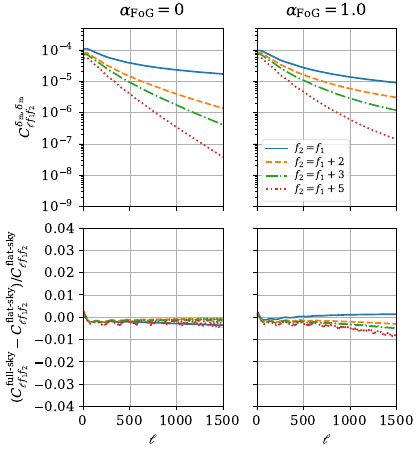}
    \caption{%
    \emph{Upper panels}: Full-sky computation of multi-frequency angular power spectrum $C_{\ell f_1 f_2}$ for the matter overdensity~$\deltam$ with $\alphaNL=1$, as described in Appendix~\ref{app:sim_details:aps}, for channel $f_1$ centered at $657.8\,{\rm MHz}$ and channel $f_2$ as shown in the legend. 
    In addition to the frequency auto-correlation $f_1=f_2$, we show frequency spacings corresponding to $k_\parallel \approx1.6\ihMpc$, $1.1\ihMpc$, and $0.64\ihMpc$.
    The left panel shows results with no Finger-of-God damping ($\alphaFoG=0$), while the right panel includes the fiducial amount of damping ($\alphaFoG=1.0$), which suppresses the angular power when $f_2=f_1$ and has a more complex effect when $f_2\neq f_1$.
    \emph{Lower panels}: Fractional difference between full-sky calculations and simpler flat-sky versions given by \cref{eq:Cell_flatsky}. The two calculations are consistent to within 1\% at the scales relevant for this work, which validates our full-sky implementation.
    }
    \label{fig:cl_comparison_matter}
\end{figure}

\begin{figure}[t]
   \centering \includegraphics[width=\linewidth, keepaspectratio, trim = 0 0 0 0]{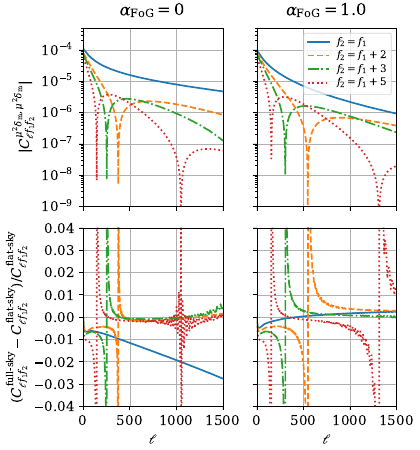}
    \caption{%
    Same as \cref{fig:cl_comparison_matter}, but for the angular power spectrum corresponding to the Kaiser term $\mu^2\deltam$. Away from zero-crossings, we find better than $3\%$ agreement between the full-sky and flat-sky calculations. This agreement can be improved by increasing the sampling of the full-sky integrand in our implementation, but we argue in the main text that this precision is acceptable for this work.
    }
    \label{fig:cl_comparison_kaiser}
\end{figure}

\begin{figure}[t]
   \centering \includegraphics[width=\linewidth, keepaspectratio, trim = 0 0 0 0]{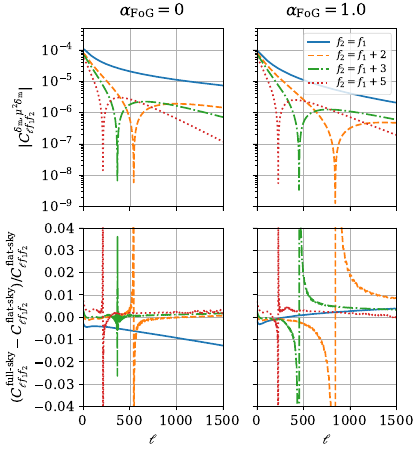}
    \caption{%
    Same as \cref{fig:cl_comparison_matter}, but for the cross power between the matter overdensity $\deltam$ and the Kaiser term $\mu^2\deltam$. We find better than $1.5\%$ agreement between the full-sky and flat-sky cases.
    }
    \label{fig:cl_comparison_matterkaisercross}
\end{figure}

In \cref{fig:cl_comparison_matter,fig:cl_comparison_kaiser,fig:cl_comparison_matterkaisercross}, we compare the full-sky and flat-sky computations of $C_{\ell f_1 f_2}$ for the matter overdensity, Kaiser term, and matter-Kaiser cross, at a representative sample of $\alphaFoG$ values and frequency separations. Ignoring regions close to zero-crossings, we find agreement to within roughly $1\%$ for $\deltam\times\deltam$, $3\%$ for $\mu^2\deltam \times \mu^2\deltam$, and $1.5\%$ for $\deltam\times \mu^2\deltam$. We have found that this can be improved by increasing the density of the $(\chi_1', \chi_2')$ grid on which $\xi_{AB}$ is sampled when evaluating \cref{eq:Clf1f2AB}.

\begin{figure}[t]
   \centering \includegraphics[width=\linewidth, keepaspectratio, trim = 0 0 0 0]{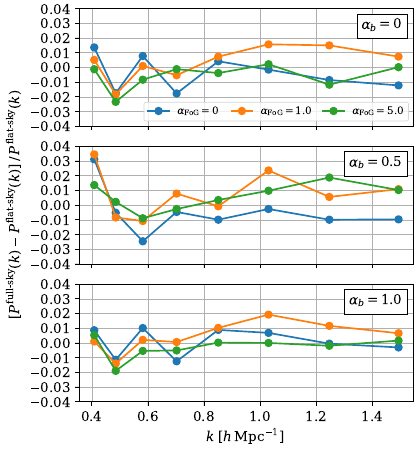}
    \caption{%
    Fractional differences between \tcm power spectra $P(k)$ measured from simulations where the input multi-frequency angular power spectra were generated in our full-sky formalism or a simpler flat-sky approximation. We show the 3 $\alpha_b$ values relevant for our template-based prediction framework, and 3 representative $\alphaFoG$ values. The largest fractional difference is roughly 3.5\%, whereas the root-mean-square fractional difference (over all $\alpha_b$, $\alphaFoG$, and $k$ values) is 1.2\%. As discussed in \secref{sec:parameter_constraints:systematics}, this is a subdominant contribution to the total uncertainty budget on the power spectrum amplitude parameter $\AHIps$.
    }
    \label{fig:pk_comparison}
\end{figure}

In \cref{fig:pk_comparison}, we compare \tcm power spectra $P(k)$ from simulations generated from full-sky and flat-sky $C_{\ell f_1 f_2}$ calculations, across the three~$\alpha_b$ values used to generate our template power spectra (see Appendix~\ref{app:model_details:template_combinations}) and three representative~$\alphaFoG$ values, for $\alphaNL=1$.
We find better than $3.5\%$ agreement in all cases, and we compute the root-mean-square fractional difference (over all $k$, $\alpha_b$, and $\alphaFoG$ values) to be $1.2\%$. This is roughly consistent with the accuracy of the template-based $P(k)$ prediction when compared with our set of 20 validation simulations, as discussed in \secref{sec:parameter_constraints:systematics:template_accuracy}.


\section{Construction of CHIME simulations based on IllustrisTNG}
\label{app:tng}

In this appendix, we provide further details about the IllustrisTNG-based CHIME simulations discussed in \secref{sec:simulations:tng}.

\subsection{Measuring {\HI} power spectra}

We follow the prescription from \cite{villaescusa-navarro2018} to post-process output snapshots from TNG100 and TNG300 to compute the {\HI} mass of each gas particle. In this prescription, for non-star-forming gas, all neutral hydrogen is assumed to be in atomic form, while for star-forming gas, the neutral hydrogen fraction and H$_2$ fractions are recomputed based on the local gas metallicity and density (see \citealt{villaescusa-navarro2018} for details). \citet{osinga2025-HImodelling} found that using other {\HI} post-processing prescriptions (e.g.\ those from \citealt{diemer2018_simpostprocessing}) only affects the {\HI} power spectrum at the few-percent level.

We add redshift-space distortions along a chosen ``line-of-sight" axis using the gas-particle velocities, and form {\HI} density grids by depositing {\HI} masses onto $1024^3$ grid points using a cloud-in-cell assignment scheme. We then measure the redshift-space {\HI} power spectrum as a function of wavenumbers parallel and perpendicular to the line-of-sight axis, in bins with boundaries $[nk_{\rm f}, (n+1)k_{\rm f})$ where $k_{\rm f} \equiv 2\pi/L_{\rm box}$ is the fundamental wavenumber of the box, and deconvolve the cloud-in-cell window function from the result. We perform these calculations using the public \texttt{Pylians} library \citep{Pylians}. We use the $z=1$ simulation snapshots, as there are no other snapshots with redshifts closer to the CHIME measurements ($z=1.16$, $1.08$, and $1.24$ for the full band and sub-bands) that contain the metadata required for {\HI} post-processing.

\subsection{Smoothing and extrapolation of {\HI} power spectra}

\begin{figure*}
   \centering \includegraphics[width=1.0\linewidth, keepaspectratio, trim = 0 10 0 0]{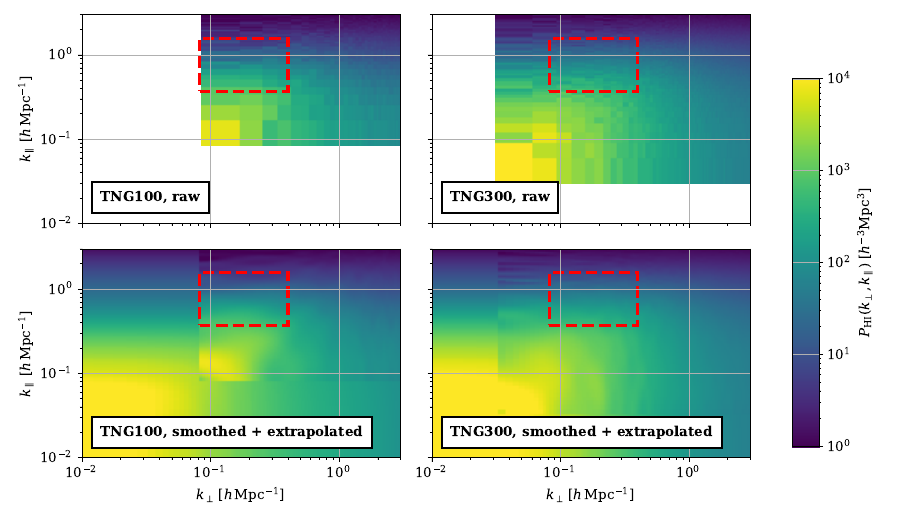}
    \caption{\emph{Upper panels:}
    Raw measurements of $P_{\sHI}(k_\perp,k_\parallel)$ from $z=1$ snapshots of the TNG100 and TNG300 runs of the IllustrisTNG simulations. The red dashed boxes indicate the range of scales probed by the CHIME power spectrum measurement. White regions denote scales that are inaccessible to each simulation run, determined by the size of each simulation box.
    \emph{Lower panels:} Versions of the power spectra from the upper panels that have been smoothed and extrapolated to lower $k$, in order to provide smooth model predictions to the CHIME simulation pipeline. The resulting CHIME simulations are used to validate the flexibility of our power spectrum model (\secref{sec:posterior_estimation:validation} and \secref{sec:parameter_constraints:tngmock}) and compare IllustrisTNG outputs to our measurements (\secref{sec:tng}).
    }
    \label{fig:tng_pk2d}
\end{figure*}

The resulting 2D {\HI} power spectra are shown in the upper panels of \cref{fig:tng_pk2d}. The scatter in the power spectra is due to sample variance in the simulations. To avoid propagating this scatter into the CHIME simulations, we fit a bivariate cubic smoothing spline to the measurements and use this smoothed power spectrum in our processing.\footnote{In detail, we fit the smoothing spline to $\log[k_\parallel^2 P_{\rm HI}(k_\parallel,k_\perp)]$, to flatten the dominant scale-dependence and reduce the dynamic range prior to smoothing. We use \texttt{scipy.interpolate.RectBivariateSpline} for this step, with smoothing factor $s=2000$ for TNG100 and $s=2250$ for TNG300, chosen by hand to reduce scatter from sample variance without eliminating physical power spectrum features.}
We verify that this smoothing does not bias the amplitude of the power spectrum by computing the ratio of smoothed and unsmoothed spectra; over the scales shown in \cref{fig:tng_pk2d}, the mean of this ratio is $1.007$ for TNG100 and $1.003$ for TNG300 .

The red dashed boxes in \cref{fig:tng_pk2d} denote the scales of the CHIME measurement: $0.37\ihMpc < k_\parallel < 1.58\ihMpc$ and $0.082\ihMpc < k_\perp < 0.40 \ihMpc$. The finite box size of each simulation sets a lower limit on the wavenumbers that can be measured along each axis, with $k_{\rm min}=0.084\ihMpc$ for TNG100 and $0.031\ihMpc$ for TNG300.\footnote{Due to the binning scheme used for the measurement of $P_{\sHI}(k_\perp,k_\parallel)$, described in the main text, modes with $k_\perp$ or $k_\parallel$ equal to $k_{\rm min}$ lie at the lower edge of their corresponding bins. This why the white (unmeasured) regions in the upper panels of  
\cref{fig:tng_pk2d} exactly correspond to the $k_{\rm min}$ values for each box.}
However, the CHIME simulation pipeline requires full-sky multifrequency HEALpix maps as inputs. The minimum $k_\perp$ and $k_\parallel$ values accessible to these maps are both lower than the minimums above, so we must choose how to fill in this lower-$k$ information.

To verify that this choice does not propagate into the CHIME simulation outputs at the scales of interest, we apply the simulation pipeline to two versions of the smoothed TNG power spectra. The first version sets the TNG power spectra to zero at wavenumbers lower than the $k_{\rm min}$ for each box (represented by the white regions in the upper panels of \cref{fig:tng_pk2d}). In the second version, we use the following power spectrum model at the missing low wavenumbers:
\begin{align*}
P_{\sHI}(k, \mu) 
	&=
	\left[ \bHI + f \mu^2 \right]^2
	P_{\rm m}(k)
	\DFoG{\sHI}(k\mu)^2
	 \\
&\quad 
	+ P_{\rm SN} 
	\DFoGSN{\sHI}(k\mu)^2\ ,
\numberthis
\label{eq:P21-theory-withSN}
\end{align*}
where $f$ and $P_{\rm m}$ are evaluated at $z=1$, $P_{\rm m}$ is given by \cref{eq:Pm-model}, $P_{\rm SN}$ is a constant, and $\DFoGSN{\sHI}$ has the same form as $\DFoG{\sHI}$ from \cref{eq:DFoGk}, but with a distinct $\sigma_{\rm FoG}$ parameter. 
The purpose of this model is to smoothly extrapolate the measured TNG power spectrum to lower $k$, and we find that this extrapolation is smoother when we include the extra term in \cref{eq:P21-theory-withSN} that is not included in our main model from \cref{eq:P21-theory}.
This model has 5 free parameters ($\bHI$, $\alphaNL$, $\sigma_{\rm FoG}$, $P_{\rm SN}$, and $\sigma_{\rm FoG,SN}$). We fit these parameters to the smoothed TNG power spectra, using measurements with $k_{\rm f} \leq k_\parallel \leq 3k_{\rm f}, \, k_{\rm f} \leq k_\perp \leq k_{\rm Ny} / 2^{3/2}$ or $k_{\rm f} \leq k_\perp \leq 3k_{\rm f}, \, k_{\rm f} \leq k_\parallel \leq k_{\rm Ny} / 2^{3/2}$, where $k_{\rm Ny} \equiv \pi N_{\rm grid}/L_{\rm box}$ is the Nyquist scale of the density grid we use. The resulting smoothed, extrapolated TNG power spectra are shown in the lower panels of \cref{fig:tng_pk2d}. We use these extrapolated versions as inputs to our CHIME simulations, but note that the final results differ by no more than 2\% if the un-extrapolated versions are used instead.

\subsection{Propagating through CHIME simulation pipeline}

Next, we convert $P_{\sHI}(k_\parallel,k_\perp)$ from the previous step into a multi-distance angular power spectrum $C_\ell(\chi,\chi')$ using \cref{eq:Cell_flatsky}.
In this calculation, we do not attempt to incorporate redshift evolution within the CHIME band, which is a reasonable approximation given the width of the band and the S/N of the measurement.
We then follow the relevant steps from \secref{sec:simulations:method}: integrating $C_\ell(\chi,\chi')$ over CHIME frequency-channel profiles, generating Gaussian HEALPix maps of $\deltaHI$, transforming these maps into visibilities, and processing these visibilities into ringmaps and then into isotropically-averaged power spectra.

\subsection{Rescaling in redshift}
\label{app:tng:rescaling}

The procedure above produces a prediction for the {\HI} power spectrum measured within the CHIME band, based on power spectra measured from the $z=1$ snapshots of TNG100 and TNG300. To convert this into a prediction for the \tcm power spectrum at the mean redshift of the band, $z=1.16$, we perform the following rescalings:
\begin{enumerate}
\item To approximately rescale the {\HI} power spectrum from $z=1$ to $z=1.16$, we multiply by the appropriate ratio of squared linear growth factors, $D^+(z=1.16)^2/D^+(z=1)^2 = 0.87$. This scaling is motivated by the expected growth of the power spectrum at linear scales, and one might expect that it is inaccurate at the nonlinear scales of our measurement. However, we have attempted to empirically estimate the appropriate scaling by comparing the $z=1$ and $z=1.15$ snapshots of each TNG run. While the $z=1.15$ snapshots do not retain the information needed to compute the {\HI} power spectrum, one can measure their total-gas power spectrum, and we compare this power spectrum at the two redshifts as a proxy for the redshift evolution of the {\HI} spectra. For both TNG100 and TNG300, we find that the ratio of redshift-space total-gas power spectra at $z=1.15$ and $z=1$ is approximately scale-independent with a mean value of $0.88$, which justifies our usage of the growth-factor ratio above.
\item We multiply the power spectrum by $\Tbar(z=1.16)^2$, with $\Tbar(z)$ given by \cref{eq:Tbarcompact}. For $\OmegaHI(z)$, we first measure $\OmegaHI$ from the $z=1$ TNG snapshots, which yields $\OmegaHI=6.35\times 10^{-4}$ for TNG100 and $5.97\times 10^{-4}$ for TNG300. We then rescale these values from $z=1$ to $z=1.16$ using the redshift-dependence of the fitting function from \cite{crighton2015}, given in \cref{eq:OmegaHIfid}; this involves a multiplication by $1.047$, which gives $\OmegaHI(z=1.16)=6.65\times 10^{-4}$ for TNG100 and $6.25\times 10^{-4}$ for TNG300.
\end{enumerate}

The combined effect of the two rescalings above is equivalent to multiplying the {\HI} power spectrum by $\Tbar(z=1)^2$, and then multiplying by $1.08$. Therefore, any potential uncertainties in the difference between TNG outputs at $z=1$ and $z=1.16$ are unlikely to be large enough to change the conclusions of \secref{sec:tng} in the comparison with CHIME's measured power spectrum.

\bibliography{paper}{}
\bibliographystyle{aasjournal}

\end{document}